\documentclass[12pt]{article}
\usepackage{amsmath}
\usepackage{graphicx}
\usepackage{natbib}
\usepackage{url} 
\usepackage{subcaption}

\newcommand{\mE}{\mathcal E}
\newcommand{\tr}{\mathrm{tr}}

\newcommand{\Ev}{\mathrm{E}}

\newcommand{\ind}{\perp \!\!\! \perp}
\newcommand{\cov}{\mathrm{cov}}

\usepackage{bm}
\usepackage{amsfonts}
\usepackage{hyperref}

\def\spacingset#1{\renewcommand{\baselinestretch}{#1}\small\normalsize} \spacingset{1}

{
  \title{\bf Functional Gaussian Graphical Regression Models For Air Quality Data}
  \author{Rita Fici\hspace{.2cm}\\
    University of Palermo, dSEAS, Palermo, Italy\\
    Gianluca Sottile\hspace{.2cm}\\
    University of Palermo, dSEAS, Palermo, Italy\\    
    Luigi Augugliaro\hspace{.2cm}\\
    University of Palermo, dSEAS, Palermo, Italy\\
    and \\
    Ernst-Jan Camiel Wit \\
    Computing Institute,  Università della Svizzera Italiana}
} 

\begin{document}

\maketitle

\begin{abstract}
Functional data describe a wide range of processes, such as growth curves and spectral absorption. 
In this study, we analyze air pollution data from the In-service Aircraft for a Global Observing System, focusing on the spatial interactions among chemicals in the atmosphere and their dependence on meteorological conditions. 
This requires functional regression, where both response and covariates are functional objects evolving over the troposphere. 
Evaluating both the functional relatedness between the response and covariates and the relatedness of a multivariate response function can be challenging. 

We propose a solution to these challenges by introducing a functional Gaussian graphical regression model, extending conditional Gaussian graphical models to partially separable functions. 
To estimate the model, we propose a doubly-penalized estimator. 
Additionally, we present a novel adaptation of Kullback-Leibler cross-validation tailored for graph estimators which accounts for precision and regression matrices when the population presents one or more sub-groups, named joint Kullback-Leibler cross-validation. 
Evaluation of model performance is done in terms of Kullback-Leibler divergence and graph recovery power.  
\end{abstract}

\noindent%
\vfill

\newpage
\spacingset{1.75} 
\section{Introduction}
\label{sec:intro}
The study of air, ocean, and land pollution often involves analyzing large datasets where observations are irregularly distributed across space. 
Air pollution disperses horizontally and vertically, affecting air quality at various elevations. 
The vertical movement of chemical elements in the atmosphere plays a crucial role in determining local meteorological conditions and air quality. 
As highlighted by \cite{JiEtAl_21}, it is essential to monitor the tropospheric vertical column density of pollutants to fully understand their impact on air quality. 
In recent years, many studies have focused on evaluating the vertical profiles of pollutants and the meteorological factors that influence them. These factors vary with altitude, leading to different effects on air quality at various heights \citep{UnoEtAl_14, SherwoodEtAl}. 

Our goal is to develop statistical tools to analyze the interactions between these chemical elements as functions of altitude. \citep{TorresEtAl_20} was one of first to apply functional data analysis (FDA) to air quality data monitoring. Additionally, we aim to explore the influence of atmospheric conditions on these chemical elements by defining a multivariate function-on-function regression approach. 
Utilizing the framework of FDA, we seek not only to examine the intricate interactions among chemical elements and their dependencies on atmospheric conditions but also to investigate how these effects vary, smoothly and continuously, across different atmospheric altitudes. 
Furthermore, we consider the conditional dependence structure of the pollutants variables in conjunction with their dependencies on external conditions. 

Some early approaches for a canonical analysis of operators are developed by \cite{LeurgansEtAl_93} and \cite{DauxoiEtNkiet_97}. 
\cite{HeEtAl_03} identify the conditions under which the canonical correlation between curves is well-defined. They use the canonical basis functions to represent the functional variables. 
Regression models that involve both dependent and independent variables as univariate functional processes have been studied by \cite{HeEtAl_10}. 
They derive the regression parameter function in terms of the canonical components of the involved processes. 
\cite{ChiouEtAl_16} formulate a multivariate functional linear regression model by employing a vectorized expansion, 
where each element of the basis system is a vector of functions with an associated scalar. 

In FDA, the Karhunen-Loève (K-L) expansion, proposed by \cite{Karhunen_1946}, stands as the most commonly utilized representation for a curve. In a multivariate setting, this expansion can be extended in various ways. 
\cite{QiaoEtAl_JASA_19} consider functions that are concentrated in a finite-dimensional subspace, and the truncated version of the K-L expansion is employed. 
The authors define the covariance for a multidimensional functional process, considering the global domain of the variables and identifying a single correlation structure. This approach simplifies earlier approaches \citep{ZhouEtAl_2010} that considered the more complex relationships among the functions. 
The framework has been extended further by \cite{MoysidisAndLi_21} to identify subpopulations of functions sharing a significant common structure of relations. While the approach in \cite{QiaoEtAl_JASA_19} establishes important foundations for the analysis of multivariate functional variables, the truncated K-L expansion, performed one curve at a time, ignores the multivariate functional structure under examination. 
In \cite{ChiouEtAl_16}, the applied expansion gives a sequence of elements constituted by a scalar coefficient and a vector of basis, one function for each variable. 

Recently, \cite{ZapataEtAl_2022} introduced a structure termed \textit{partial} \textit{separability} for the covariance operator of multivariate functional data. This approach defines a vectorized version of the K-L expansion where the variables share a unique basis system. 
The idea of partial separability builds on earlier developments in the literature. 
The concept of \textit{separable} covariance has been applied previously to the study of random fields with a spatio-temporal domain \citep{Genton_2007, GneitingEtAl_2006}. 
\cite{DelicadoEtAl_10} describes the possible contributions that the combination of functional data analysis and geostatistics methodology can give, and shows that multivariate spatial statistical tools can be generalized to be valid for functional data. 
\cite{GromenkoEtAl_12} develop an estimation methodology for the functional mean and the functional principal components when the functions form a spatial field. 
In functional terms, the separability assumption means that the covariance structure factors into the product of two functions. 
\cite{AstonEtAl_17} and \cite{ConstantinouEtAl_18} define the concept of \textit{separability} for functional data in terms of tensor product. 
\cite{LynchEChen_18} introduce the concept of weak separability to support factorization methods that decompose the signal into its components. 
The covariance operator is parametrized by a pair of basis functions resulting from the eigenfunctions of the marginal covariance kernels. 

The notion of graphical models for multivariate functional data was proposed by~\cite{ZhuEtAl_JMLR_16}. The authors propose to construct the graphical model directly in the space of infinite dimensional random functions through establishing the Markov distributions and hyper Markov laws for random processes. In the setting of multivariate Gaussian process (MGP), the authors also propose to use a hyperinverse-Wishart process prior for the covariance kernels of the coefficient sequences, and study the theoretical properties of the proposed prior, such as existence and uniqueness. \cite{LeeEtAl_21} introduced the notion of conditional functional graphical model for functional data, but due to lack of separability, the structure is complex and computationally intensive. \cite{QiaoEtAl_BioK_20} proposed a dynamic functional graphical model that allows the graph structure to change over time. Although the approach proposed in~\cite{QiaoEtAl_JASA_19} is appealing, the authors show that their approach can recover the true edge-set only under the restrictive assumption that each random function is a finite dimensional object. As pointed out by~\cite{ZapataEtAl_2022}, the reason of this theoretical restriction lies in the observation that the covariance operator, which is the infinite-dimensional counterpart of the covariance matrix for standard multivariate Gaussian distribution, is compact and thus not invertible.  Consequently, the connection between conditional independence and an inverse covariance operator is lost, as the latter does not exist. To overcome this methodological problem, the authors introduce a specific assumption on the structure of the covariance operator called partial separability, which is an assumption about the eigenfunctions of the covariance operator. The concept of \textit{partial separability} for the covariance operator of multivariate functional data \citep{ZapataEtAl_2022} overcomes the problem of a non-invertible covariance operator. This concept is particularly suitable for the definition of functional graphical models.

A penalized approach for MGP is proposed in \cite{QiaoEtAl_JASA_19}. From a methodological point of view, their approach is structured in two steps. First, the authors introduce the notion of conditional cross-covariance function, which is a bivariate function representing covariance between two random functions conditional on the remaining random functions. This notion allows the author to extend to infinite dimensional setting the well-known relationship between edge-set and conditional covariance. Then, to estimate the edge-set the authors use the Karhunen-Loève expansion to obtain a finite representation of each random function, which is defined using the first $L$ functional principal scores. Since it is assumed that the vector of random functions follows a MGP, it is easy to show that the resulting vector of principal scores follows a multivariate Gaussian whose precision matrix has a specific block structure induced by the properties of the scores of the Karhunen-Loève expansion. To encourage blockwise sparsity in the precision matrix, the authors propose a generalization of the graphical lasso (glasso) estimator~\citep{YuanEtAl_BioK_07} called functional graphical lasso. 

\cite{ZapataEtAl_2022} show that partial separability is equivalent to the assumption that the vector of random functions admits a multivariate Karhunen-Loève expansion. Since the precision matrices associated to the functional precision scores contain all the necessary information to recover the conditional independence graph for the MGP, the authors propose an extension of the glasso estimator called joint graphical lasso (jglasso) estimator, which is a doubly-penalized likelihood estimator. \cite{LiEtAl_JASA_18} propose an additive nonparametric approach which does not rely on any distribution assumption. However, additive conditional independence is not equivalent with classical conditional independence. More recently, \cite{ZhaoEtAl_EJS_24} propose a functional generalization of the neighborhood selection method \citep{MeinshausenEtAl_AoS_06} which avoids defining the precision operator.


Combining multivariate regression with a conditional graphical analysis of the multivariate response involves a set of regression parameters, on the one hand, and a set of covariance parameters, on the other.  The problem of estimating two matrices of parameters, i.e., the regression matrix and the inverse of the conditional covariance matrix of the response variables under a 
multivariate Gaussian assumption, was first studied by \cite{RothmanEtAl_10} and then extended by \cite{YeeandLiu_12}. 
They propose to use a double penalized maximum likelihood estimator with respect to a normal distribution. 
Recently, \cite{SottileEtAl_2022} propose a conditional graphical lasso that infers interactions both within and between groups of functions. They devise a computationally efficient strategy for the penalized inference of the network of dependencies in high dimensions, both at the level of responses and covariates. 
A possible way to select the tuning parameters is to use information criteria, such as the Akaike Information Criterion (AIC) or the extended Bayesian Information Criterion (eBIC) 
\citep{Chen&Chen,FoygelEtAl_10}. \cite{WitEtAl_15} proposed an efficient Kullback-Leibler cross-validation (KLCV) approach, specifically designed for Gaussian graphical models, that has good finite sample sample properties and converges to the AIC.



In this paper, we extend the concept of conditional graphical models in a functional setting. We propose a doubly-penalized estimator 
along with an efficient algorithm for recovering both the conditional independence structure of the response functions and the effects of the predictor functions on the expected response curves. Our approach addresses the limitations of previous works by considering the multivariate and functional aspects of the data simultaneously while incorporating penalization to promote sparsity in the estimation. To evaluate the goodness of fit, we propose a modified version of the Kullback-Leibler Cross-Validation criterion tailored for model selection in the context of penalized conditional estimators. 

The remainder of this paper is organized as follows.
Section~\ref{sec:data} describes the main aspects of the IAGOS global initiative to monitor pollutants in the atmosphere and the vertical sampling strategy used.  In Section~\ref{sec:fcggm}, we present the methodological aspects of our proposed extension of functional graphical regression models. In Section~\ref{sec:estimator}, we provide an estimation procedure to recover the edge set using a conditional group lasso estimator. 
In Section~\ref{sec:goodness_of_fit}, we propose a novel selection method for the tuning parameter of the conditional graphical lasso estimators. 
Section~\ref{sec:simulation} provides computational details and evaluates our approach's performance through a simulation study. 
Finally, in Section \ref{sec:illustration}, we present the results of our analysis on the IAGOS data.

\section{IAGOS: high-resolution air quality monitoring}\label{sec:data}

In-service Aircraft for a Global Observing System (IAGOS) is a European research infrastructure that collects high-resolution atmospheric data from instruments installed on commercial aircrafts. 
These instruments record various air quality measurements, such as chemical concentrations, as well as meteorological variables and latitude, longitude and altitude. 
The data are available at \url{https://iagos.aeris-data.fr}. 
The IAGOS program operates in collaboration with multiple airlines, using long-haul passenger aircraft to perform quasi-continuous measurements of atmospheric 
conditions, including trace gases, aerosols, and cloud particles. Each aircraft is equipped with the IAGOS-CORE rack, which houses fully automated 
instruments that almost continuously measure various atmospheric components during flights. 
The data collection aims to provide comprehensive global coverage of air quality and atmospheric chemistry, offering high-resolution data for research and policy-making. 
The IAGOS-CORE system is designed to operate safely on commercial aircraft, complying with civil aviation standards set by the European Aviation Safety Agency (EASA). 
The system is installed in the fuselage, with dedicated inlet probes mounted on a specialized plate. 
This setup allows for precise atmospheric measurements without interfering with the operation of the aircraft. 
The sensors are designed to capture concentrations of key chemical species and meteorological variables at different altitudes. 
The data is recorded at high temporal resolution (4 seconds), allowing for detailed analyses of air quality dynamics over space.

\begin{figure}
  \begin{minipage}[b]{0.45\textwidth}
      \includegraphics[width=\linewidth]{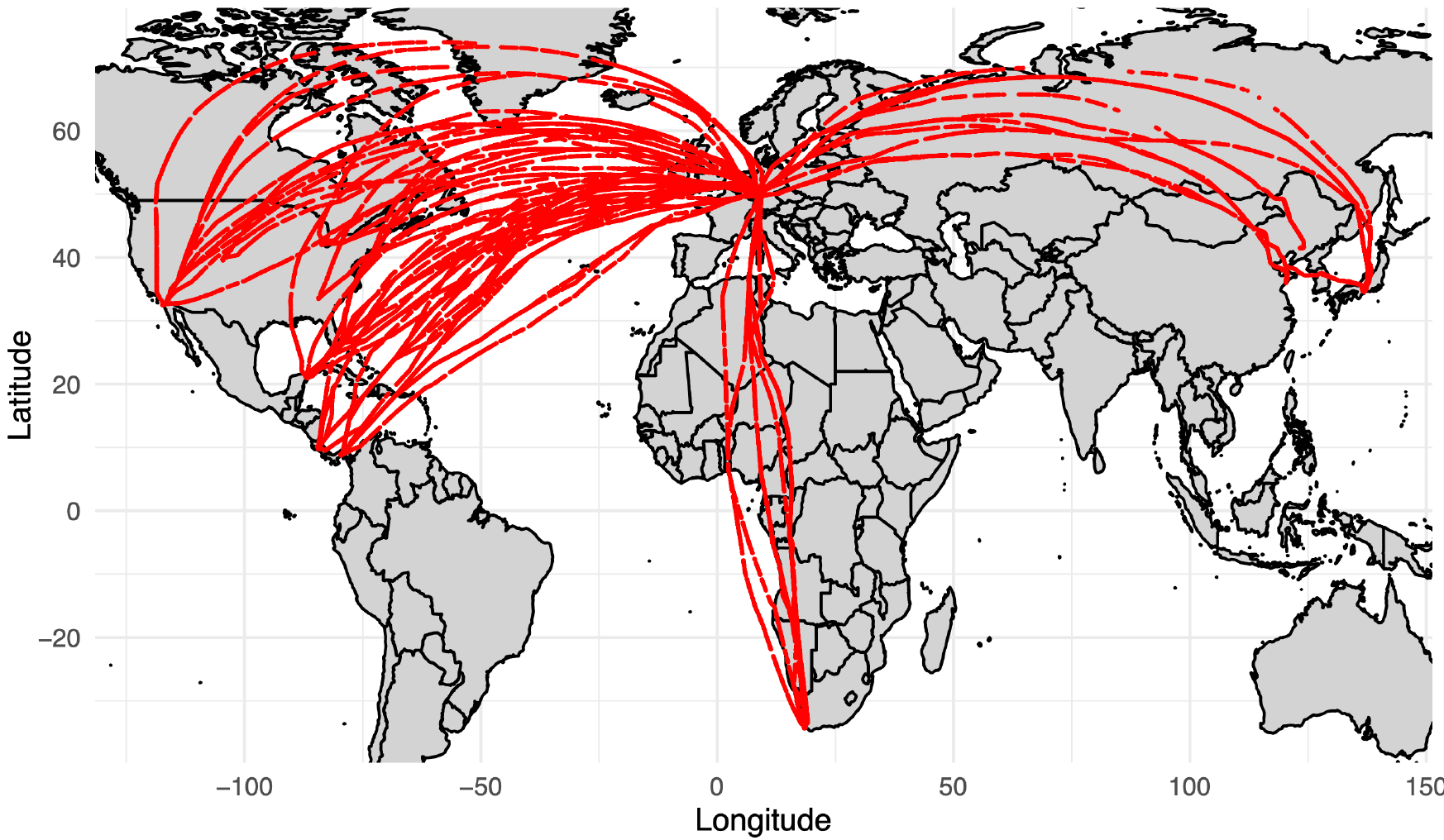}
      \caption{Flight trajectories from the IAGOS dataset.}
      \label{fig:map}
  \end{minipage}
  \hfill
  \begin{minipage}[b]{0.45\textwidth}
      \includegraphics[width=\linewidth]{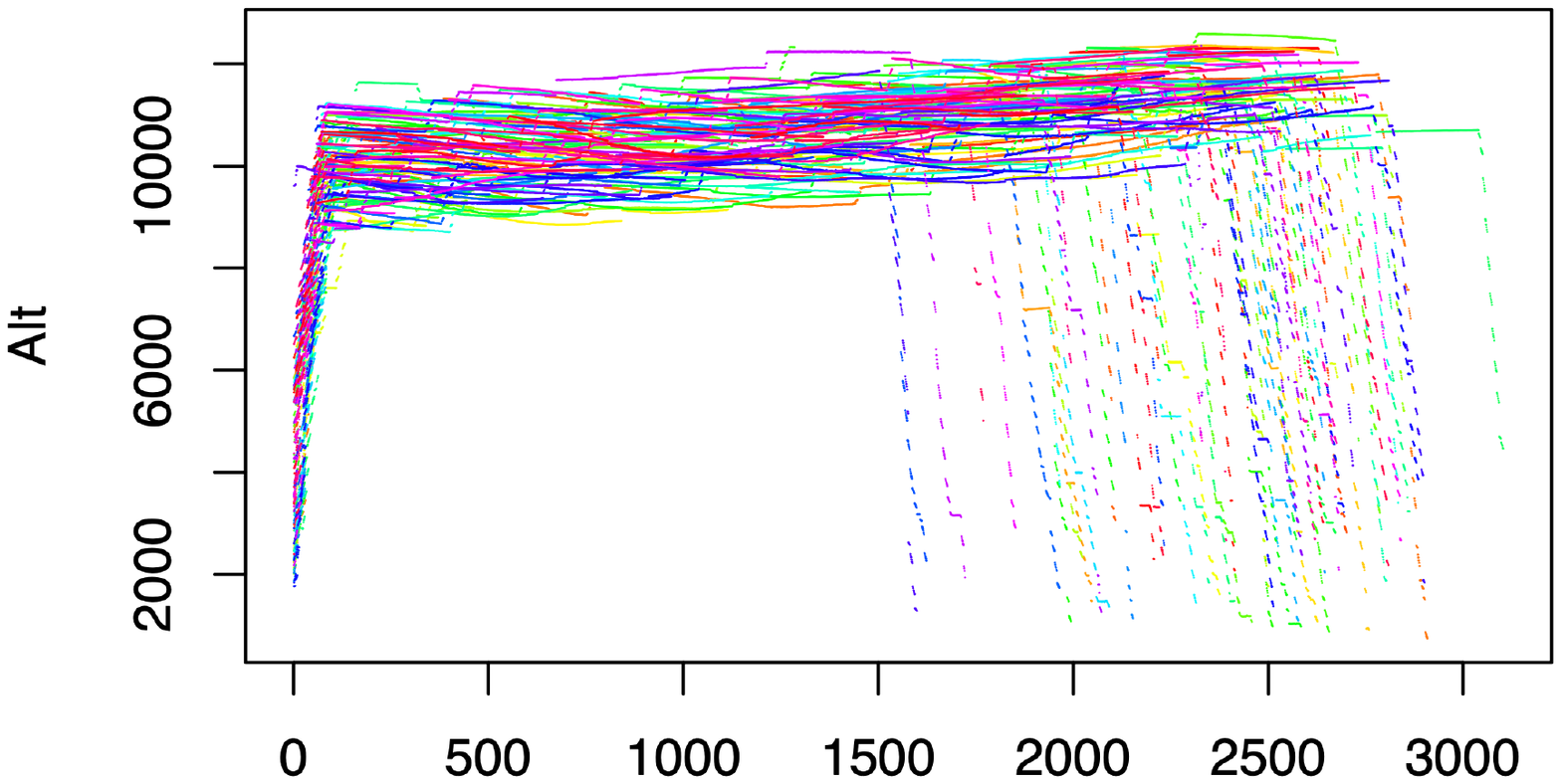}
      \caption{Observed altitudes during the flights expressed in meters.}\label{fig:alt}
  \end{minipage}
\end{figure}

\begin{figure}[!h]
  \centering
  \includegraphics[scale = 0.55]{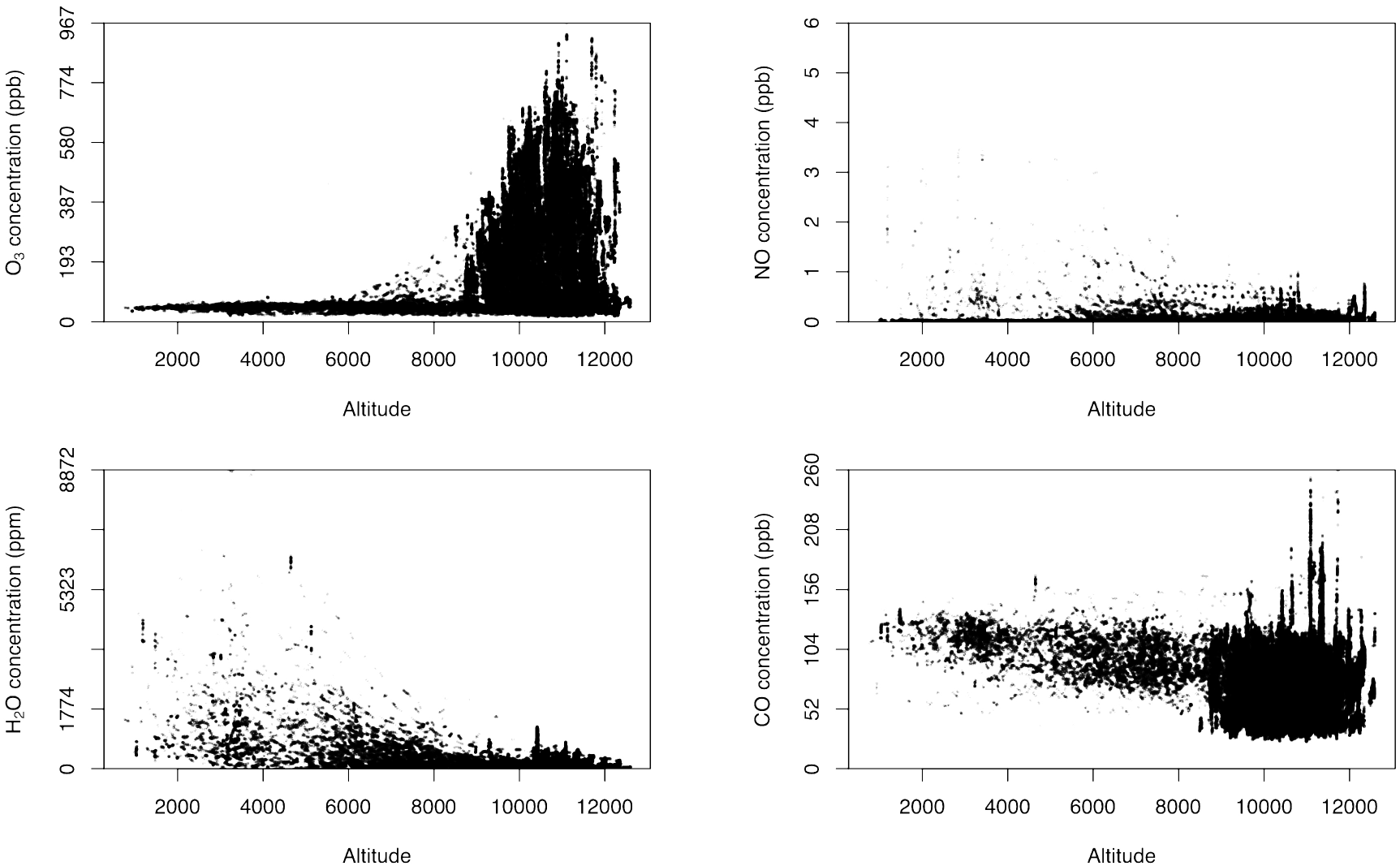}
\\ \vspace{1cm}
  \includegraphics[scale = 0.35]{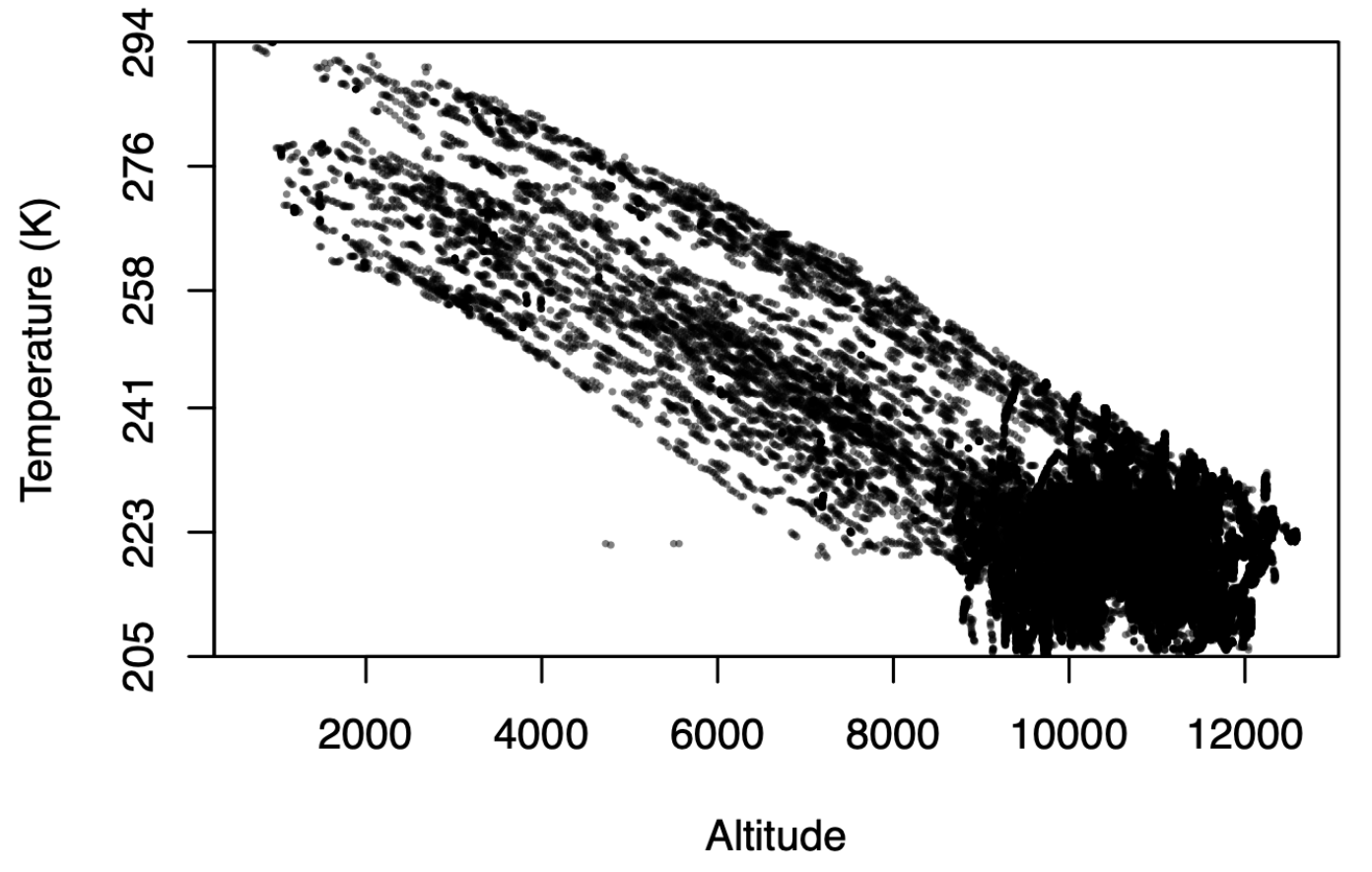}
  \caption{Observed values of chemicals concentrations and temperature over altitudes. $O_3$, $NO$, and $CO$ are measured in parts per billion (ppb), while $H_2O$ is measured in parts per million (ppm). Temperature is measured in Kelvin (K).}  \label{fig:scatter plots}
\end{figure}

We consider data from $75$ flights in 2020. Each flight contributes as a statistical unit to the dataset. 
Figures \ref{fig:map} and \ref{fig:alt} show their trajectories and altitudes. On average, each flight has 2450 observed points over the altitude. 
The original dataset includes records at altitudes up to 15000 $m$, but we analyze the data up to 13000 $m$ to maintain a very dense grid. 
We analyze five air quality dimensions: the concentrations of Carbon Monoxide ($CO$), Ozone ($O_3$), Nitrogen Monoxide ($NO$), Water vapour ($H_2O$), and the air temperature. 
$CO$ is a gas produced by incomplete combustion of fossil fuels. 
It serves as an important tracer for anthropogenic pollution and plays a significant role in forming ground-level Ozone. 
$O_3$ is a reactive gas. Its presence in the stratosphere protects the Earth from harmful ultraviolet radiation. 
At ground level, however, it is a pollutant and harms human health. 
$NO$ is a precursor to Ozone formation and an indicator of anthropogenic pollution, particularly from vehicle emissions and industrial activities. 
$H_2O$ is a key component influencing both weather patterns and the transport of pollutants, with its concentration varying significantly with altitude. 
The distribution of the chemicals over the altitude differs. 
As shown in Figure \eqref{fig:scatter plots}, the concentrations of $CO$ and $O_3$ reach their highest values at elevated altitudes, 
while the concentrations of $NO$ and $H_2O$ exhibit the opposite trend. The concentration of points on the right side of each plot in Figure \eqref{fig:scatter plots} 
reflects the denser grid of records at higher elevations, a pattern also evident in Figure \eqref{fig:alt}. 
The mean distance between two consecutive observations is 8 meters.

\section{Functional Gaussian graphical regression}
\label{sec:fcggm}
In our study, the response random functions, denoted by $\bm Y$, represent the vertical concentrations of $O_3,\, CO,\, NO$, and $H_2O$. 
The set of covariates, $\bm X$, consists of a single variable: the vertical profile of temperature. 
For the sake of generality in the developed methodology, we will maintain the vector notation for both $\bm Y$ and $\bm X$. 
Measurements of chemical concentrations and temperature are taken over a discrete and irregularly spaced 
grid of altitudes, which varies for each statistical unit. 
FDA is particularly suitable for modeling phenomena where variables evolve smoothly across continuous dimensions, 
providing a flexible framework for capturing complex relationships between environmental factors and spatial features.

Let us consider two multivariate processes with functional components $\bm Y = \{\bm{Y}(s) \in \mathbb R^p: s \in \mathcal S\}$ 
and $\bm X = \{\bm{X}(s) \in \mathbb R^q: s \in \mathcal S\}$ 
observed over a closed and continuous interval of the altitude, $\mathcal{S} \subset \mathbb R$. 
The processes belong to the Hilbert spaces, denoted by $\mathbb H_{\bm Y}$ and $\mathbb H_{\bm X}$, 
of $p$- and $q$-dimensional vectors of square-integrable functions $\mathcal L_2^p[\mathcal S]$ and $\mathcal L_2^q[\mathcal S]$.

The vectors of observations over the altitude grids can be understood as discrete versions of 
realizations of underlying smooth functions, which are affected by an error term (see \cite{RamsaySilverman_05}),
$$Y_j^o(s)=Y_{j}(s) + e^y_{j}(s) \text{ and } X_k^o(s)=X_{k}(s)+ e^x_{k}(s) \text{ for }j = 1, \dots, p \text{ and }\,j = k, \dots, q\,;$$
where $ Y_{j}^0(s) \text{ and } X_{k}^o(s)$ are the observed realizations of noisy versions of $Y_j \text{ and } X_k$  at a discrete altitude $s$,
and $ e^y_{j}(s) \text{ and } e^x_{k}(s)$ are measurement errors. 
There is a significant amount of measurement error in air-data recording, so the functional variables are not directly observed. 
The remaining part of this section details modelling the smooth functions $\bm Y$ and $\bm X$, 
while the estimation of the error terms is described in Section~\eqref{sec:estimator_score-stim}.

\subsection{Multivariate Functional Regression Model}
In this section, we define a functional regression model where the sum of two components gives the response process: 
\begin{equation}
\label{eq:Ymodel}
\bm Y(t)= \int_{\mathcal S} \bm \beta(t,s) \bm X(s) ds+\bm K(t)\,, \,\,\, t \text{ and }s \in \mathcal{S}.
\end{equation}
The first element on the right-hand side of \eqref{eq:Ymodel} is a structural component that considers the effects of the covariates on the response. 
It involves the covariate process $\bm X$ and a set of bivariate regression coefficient functions $\bm\beta(t,s)\in\mathcal L_2^{p \times q}[\mathcal S\times \mathcal S]$, 
which represent how the effect of $\bm X$ on $\bm Y$ changes over $\mathcal{S} \times \mathcal{S}$. 
If the $\mathit{L}_2$ norm of $\beta_{ik}$ vanishes, 
$\|\beta_{ik}\| = 0$, then $Y_i$ is independent of $X_k$, given $\bm X_{- k}$. 
If $\|\beta_{ik}\| \not = 0$, then $X_k$ is a predictor of $Y_i$.
The term $\bm K(t)$ represents the noise component, independent of $\bm X$, and encodes the mutual dependencies among the responses, net of the effect of $\bm X$.

The two elements on the right-hand side of~\eqref{eq:Ymodel} are associated with a set of directed links and a set of undirected links, respectively. 
The first set represents the conditional dependencies of the responses on the covariates and is defined as
$$\mathcal{\mE}^{\bm X \bm Y} = \{(i,k): Y_i \not\ind X_k \mid \bm Y_{- i}, \bm X_{- k} \}.$$ 
The second set represents the pairwise conditional independence structure of $\bm Y$, and is given by
$$\mathcal{\mE}^{\bm Y} =\{ (i,j): Y_i \not\ind Y_j \mid \bm Y_{- ij}, \bm X \}= \{(i,j): K_i \not\ind K_j \mid \bm K_{- ij} \}.$$ 
Note that conditional independence holds if the functions are conditionally independent over the entire domain $\mathcal{S}\times \mathcal{S}$. 
These sets of edges define a chain graphical model with two components denoted by $\mathcal{G}= \{ \mathcal{V}, \mathcal{\mE}^{\bm X \bm Y}, \mathcal{\mE}^{\bm Y} \}$, 
where $\mathcal{V}$ has cardinality $p+q$. 
The directed edges in $\mathcal{\mE}^{\bm X \bm Y}$ correspond to the first level factorization of the probability density $f_{XY}$, which for two chain components is 
\begin{equation}\label{eq:first_fac}
f\left(  \bm X,\bm Y \right) = f_X(\bm X) f_Y (\bm Y \mid \bm X_{pa})
 \end{equation}
where $\bm X_{pa}$ denotes the set of elements of $\bm X$ that affect $\bm Y$. 
The undirected links in $ \mathcal{\mE}^{\bm Y}$ correspond to the second-level factorization 
\begin{equation}\label{eq:second_fac}
    f(\bm Y \mid \bm X_{pa}) = \prod_{C\in \mbox{\scriptsize Cliques}} f_C(\bm Y_C \mid \bm X_{pa})
\end{equation}
where $C$ are the cliques in $\mathcal{\mE}^{\bm Y}$. 
Although in this analysis, we are not directly interested in the factorization of $f_X(\bm X)$, the proposed framework can be extended to include a graphical description of $\bm X$. We assume here that $\bm X$ and $\bm K$ are multivariate Gaussian processes with mean zero.

\subsection{Auto-covariance operator}
To define the dependence structure on the Hilbert space and extend the definition of edge set related to a single graph, we consider the correlation among the functions across the entire domain, as suggested by \cite{QiaoEtAl_JASA_19}. 
For the vector of response variables, the covariance is defined as a matrix $\mathcal{C}$ where each entry involves an integral operator with a symmetric and nonnegative definite kernel equal to
\begin{equation}\label{cond.cov.fun}
C_{ij}(t,t^\prime)= \cov\left(Y_i \left(t \right), Y_j \left(t^\prime\right) \right)\,, \hspace{0.2cm} t \text{ and } t^\prime \in \mathcal S.
\end{equation}
The auto-covariance operator of the process is defined as
$$(\mathcal C \bm Y)(t) = \int_{\mathcal S} C(t,t^\prime) \bm Y(t^\prime) dt^\prime \,$$
where $C(t,t^\prime)$ is the $p \times p$ symmetric matrix which collects the kernels \eqref{cond.cov.fun} for $i,j = 1,\dots, p$.
If $\mathcal C$ is \textbf{partially separable} ( \cite{ZapataEtAl_2022}) there exists a sequence of $p\times p$ dimensional matrices, $\{\Sigma_l\}_{l=1}^{\infty}$, and a set of orthonormal eigenbasis $\{\varphi_l\}_{l=1}^{\infty}$, such that it is possible to rewrite $\mathcal C$ as
\begin{equation}
\label{expansionC}
 \mathcal C = \sum^{\infty}_{l=1}\Sigma_{l}\varphi_l \otimes \varphi_l
\end{equation}
where $\otimes$ is the tensor product and $\{\varphi_l\}_l^{\infty}$ is a set of orthonormal functions in $\mathcal S$. The trace of $(\Sigma_l)$ decreases over $l$. 
Thanks to~\eqref{expansionC}, it is possible to represent the multivariate process as,
\begin{equation}
\label{vectorizedKL1}
\pmb Y(t)= \bm\mu(t) + \sum^\infty_{l=1}\bm{\gamma}_l\varphi_l(t)\, ,
\end{equation}
where $\bm\gamma_l$ is a vector of scores with dependent entries. 
The dependence structure of the process is directly related to the dependence among the scores inside each vector $\bm\gamma_l$, i.e., in general $\gamma_{il}\not\ind\gamma_{jl}$. 
However, the sequence of vectors of scores are independent of each other, i.e., $\bm \gamma_l \ind \bm \gamma_{l^\prime}$ for $l \not = l^\prime$. 

A fundamental advantage of~\eqref{expansionC} is that, if the scores are normally distributed, the elements in $\{\Sigma_l\}_{l=1}^\infty$ identify the pairwise relations among the variables in the process.
Due to the independence of the vectors from each other, these covariance relations can be visualized as a block diagonal infinite-dimensional covariance matrix, denoted by $\bm\Sigma$, easy to invert,
$$
\bm\Sigma =  \begin{bmatrix} \Sigma_{1} & 0 & \dots \\
0 & \Sigma_{2} &\dots\\
\vdots & \vdots &\ddots  \\
\end{bmatrix} \text{ and } \Theta = \bm\Sigma^{-1} = 
 \begin{bmatrix} \Sigma^{-1}_{1} & 0 & \dots \\
0 & \Sigma^{-1}_{2} &\dots\\
\vdots & \vdots &\ddots  \\
\end{bmatrix}.
$$
In both matrices, the off-diagonal block entries are equal to zero. 
In this paper, we exploit the advantage of representing some variables in a unique system of orthonormal basis by applying the assumption of partial separability 
on the autocovariance operator of $\bm Y$, in $\mathbb{H}_{\bm Y} \times \mathbb{H}_{\bm Y}$, and on the autocovariance operator of $\bm X$, in $\mathbb{H}_{\bm X} \times \mathbb{H}_{\bm X}$. 
The processes are represented as,
\begin{equation}\label{exp.YX}
  \bm Y(t) = \sum^\infty_{l=1} \bm\gamma_l\varphi_l(t) \vspace{1cm}\text{ and } \vspace{1cm} \bm X(s) = \sum^\infty_{h=1} \bm\chi_h\psi_h(s).
\end{equation}

\subsection{Connection of parametrization with graphical structure\label{subsec:graph_structure}}

\paragraph{Identification of covariate-response conditional independence graph.}
In factorization \eqref{eq:first_fac}, the set $\bm X_{pa}$ is identified by the regression functions such that $X_k$ is a parent of $Y_i$ if and only if
$\| \beta_{ik} \| \not= 0$. In this section, we derive a representation of the parental set through the normally distributed score vectors $\bm\gamma_l$ and $\bm\chi_h$ in \eqref{exp.YX}.
\vskip .1in
\noindent
\textbf{Theorem 1.}\textit{ Let $\bm Y$ and $\bm X$ be two multivariate Gaussian functional processes, their regression dependence can be represented by the coefficient functions $\bm\beta$ which satisfy three characterizations:
\begin{enumerate}
\item the linear model 
\begin{equation}
\label{statement1}
\bm Y (t)= (\mathcal B \bm\beta_0)(t) + \bm K(t)
\end{equation} where $ \Ev[\|\bm Y (t)\|_{\mathbb H_{\bm Y}} \mid \bm X]  < \infty$ and $\mathcal{B}: \mathbb H_X \times \mathbb H_Y \rightarrow \mathbb H_Y$ is the regression operator defined as 
$$
(\mathcal B \bm \beta)(t)=\int_{\mathcal S}\bm \beta(t,s) \bm X(s)ds \, ;
$$
\item the population normal equation 
\begin{equation}
\label{statement2}
\cov[ \bm Y(t), \bm X(s)]=(\mathcal D_{\bm X\bm X} \bm \beta_0) (t,s)\end{equation}
 where $\mathcal D_{\bm X\bm X}:\mathbb H_{\bm X} \times \mathbb H_{\bm X} \rightarrow \mathbb H_{\bm Y} \times \mathbb H_{\bm X}$ is the cross-covariance operator defined as,
 $$
 (\mathcal D_{\bm X\bm X}\bm \beta)(t,s)=\int_{\mathcal S} \bm \beta(t,s)\cov[\bm X(s), \bm X(s^\prime)]ds^\prime;
 $$
 where $s^\prime$ is a point in $\mathcal S$;
\item  and the optimisation on the squared norm of the difference between $\bm Y$ and $(\mathcal B \bm\beta)$ in the response Hilbert space,
\begin{equation} 
\label{statement3}
\bm\beta_0 = \underset{\beta}{\operatorname{argmin}}\text{ } \Ev\| \bm Y - (\mathcal B \bm\beta) \|^2_{\mathbb H_{\bm Y}}.\end{equation}
\end{enumerate}}
\vskip .1in 
\noindent
Theorem 1 follows from the generalization of Proposition 2.2 in \cite{HeEtAl_10}. In \cite{HeEtAl_10}, the authors define the 
linear regression coefficient between two functions. Theorem 1 defines a linear multivariate regression parameter with multiple covariates. 
As the Hilbert-Schmidt operators $\mathcal B$ and $\mathcal D_{\bm X\bm X}$ do not have bounded inverses, we use the vectorized K-L expansion to extend the results 
of Theorem 2.3 in \cite{HeEtAl_10} to the multivariate case. This allows us to derive a unique solution that satisfies \eqref{statement1}, \eqref{statement2} and  \eqref{statement3}.

Considering the expansions \eqref{exp.YX} with 
$
\sum^{\infty}_{l,h=1}\| \Ev [\bm\gamma_l\bm\chi_h^\top] \Ev [\bm\chi_h\bm\chi_h^\top]^{-1}\|^2<\infty,
$ we now use Theorem 1 to characterize the regression functions as 
\begin{equation}
    \label{beta.D}
\bm \beta(t,s)=(\mathcal D_{\bm X\bm X}^{-1} \cov[ \bm Y, \bm X])(t,s) \, ,
\end{equation}
where $\mathcal D^{-1}_{\bm X \bm X}$ exists for a subdomain of $\mathcal D_{\bm X \bm X}$ and takes the form
\begin{equation}\label{subdomainD}
(\mathcal D^{-1}_{\bm X \bm X} \cdot)(t,s)= \sum^\infty_{h,l=1}  \langle\cdot, \bm\varphi_l \otimes \bm\psi_h \rangle \Ev [\bm\chi_h\bm\chi_h^\top]^{-1} \odot \left(\bm\varphi_l(t)\bm\psi_h^\top(s) \right)\, ,
\end{equation}
where $\odot$ is the Hadamard product and $\{\bm\varphi_l \otimes \bm\psi_h\}_{l,h=1}^{\infty}$ is a orthonormal basis system in $\mathbb H_{\bm Y}\times \mathbb H_{\bm X}$. 
As we apply the vectorized K-L expansion, $\bm\varphi_l$ and $\bm\psi_h$ represent sets of p- and q-dimensional functions, respectively, where all functions within each set are identical. 
Plugging in the covariance functions into $\langle\cdot, \bm\varphi_l \otimes \bm\psi_h \rangle$ and using the definition of the inner product given in Section \eqref{subsec:stcha_fun_spaces}, we have
\begin{equation}
\label{inner.cov}
\int_{\mathcal{S}}\int_{\mathcal{S}} \cov[\bm Y(t),\bm X(s)]\odot  \left(\bm\varphi_l(t)\bm\psi_h^\top(s) \right) dsdt = \textrm E[\bm\gamma_{l}\bm\chi_{h}^{\top}] \, .
\end{equation}
Thanks to \eqref{subdomainD} and \eqref{inner.cov}, we can write \eqref{beta.D} as
\begin{equation}
\label{beta_exp}
\bm \beta (t,s) =\sum\limits^{\infty}_{h,l=1}  \textrm E[\bm\gamma_{l}\bm\chi_{h}^{\top}]\textrm E[\bm\chi_h\bm\chi_h^\top]^{-1}\odot \left(\bm\varphi_{l}(t)\bm\psi^\top_h(s)\right)\, .
\end{equation}
It follows that the conditional expectation of $\bm Y(t)$ can be written as
$$
\textsc E[\bm Y(t)\mid \bm X] = \int_{\mathcal{S}}\bm \beta (t,s)\bm X(s)  ds = \sum\limits_{h,l=1}^{\infty}\bm B_{hl} \bm \chi_h \varphi_l (t)
$$
where $\bm B_{hl}=\textrm E[ \bm\gamma_{l}\bm\chi_{h}^{\top}]\textrm E[ \bm\chi_h\bm\chi_h^\top]^{-1}$. Therefore, we have now shown that the edge set $\mathcal{\mE}^{\bm X \bm Y}$ is characterized by
$$\mathcal\mE^{\bm X \bm Y} = \bigg\{(i,k)~\mid ~ \exists \, l, h: b_{hl\,ik} \not=0 \bigg\} \, ,$$
where $b_{lh\,ik}$ is the $ik^{th}$ entry of $\bm B_{hl}$.

\paragraph{Identification of response conditional independence graph.}
Given that the conditional distribution $f_Y( \bm Y\mid \bm X_{pa})$ is a multivariate Gaussian process the conditional independence structure can be expressed through the zero pattern on the inverse covariance operator. 
The covariance of $\bm Y$ given $\bm X$ is equal to the covariance of $\bm K$, and it has kernel equal to
$$ \bm C^{\bm Y \mid \bm X}(t, t^\prime) = \left\{ \cov\left(Y_i\left(t\right), Y_j\left(t^\prime\right)\mid  \bm X\left(\cdot\right)\right)\right\}_{i,j=1}^{p} =  \left\{ \cov\left(K_i\left(t\right), K_j\left(t^\prime\right)\right)\right\}_{i,j=1}^p = \bm C^{ \bm K }(t, t^\prime) \, .$$  
Thanks to Theorem~3 in~\cite{ZapataEtAl_2022}, we have the partial correlation between $K_i$ and $K_j$ related to the precision matrices of the socres $\{\bm\Theta^\gamma_l \}^\infty_{l=1}= \{(\bm\Sigma_l^\gamma)^{-1}\}_{l=1}^\infty$
$$\cov(K_i(t), K_j(t^\prime) \mid \bm K_{-ij}(\cdot))=\sum_{l = 1}^{\infty} \cov(\gamma_{l\,i}, \gamma_{l\,j}\mid \bm\gamma_{l\,-ij} ) \varphi_l (t) \varphi_l(t^\prime) = - \sum_{l = 1}^{\infty} \frac{\theta_{l\,ij} \varphi_l(t)\varphi_l(t^\prime)}{\theta_{l\,ii} \theta_{l\,jj} - \theta^2_{l\,ij}}
\, ,$$
where $\theta_{l\,ij}$ is an entry of $\bm\Theta^{\bm\gamma}_l$, which, thanks to the K-L expansion, does not depend on $\mathcal{S}$.
We can identify $\mathcal{E}_{\bm Y}$ as
$$\mathcal{\mE}^{\bm Y} = \left\{(i\,j) ~\mid~ \exists l:  \theta^{\bm \gamma}_{l\,ij} \not =0 \right\}\, .$$


\section{Functional Gaussian graphical regression estimation} 
\label{sec:estimator}
We observe $N$ independent observations of the noisy curves $\{\bm y_n\}_{n=1}^N$ and $\{\bm x_n\}_{n=1}^n$ at discrete locations $s\in\mathcal{S}$. In section~\ref{sec:estimator_score-stim}, we estimate the underlying true smooth realizations $\bm X$ and $\bm Y$ of the functional variables. Then in section~\ref{sec:estimator_graph-stim} we derive an estimator of relevant functional predictors $\bm X$, as well as the conditional dependence structure on $\bm Y$. 

\subsection{Evaluation of the scores~\label{sec:estimator_score-stim}}
\begin{figure}[!ht]
  \centering
  \includegraphics[scale = 0.7]{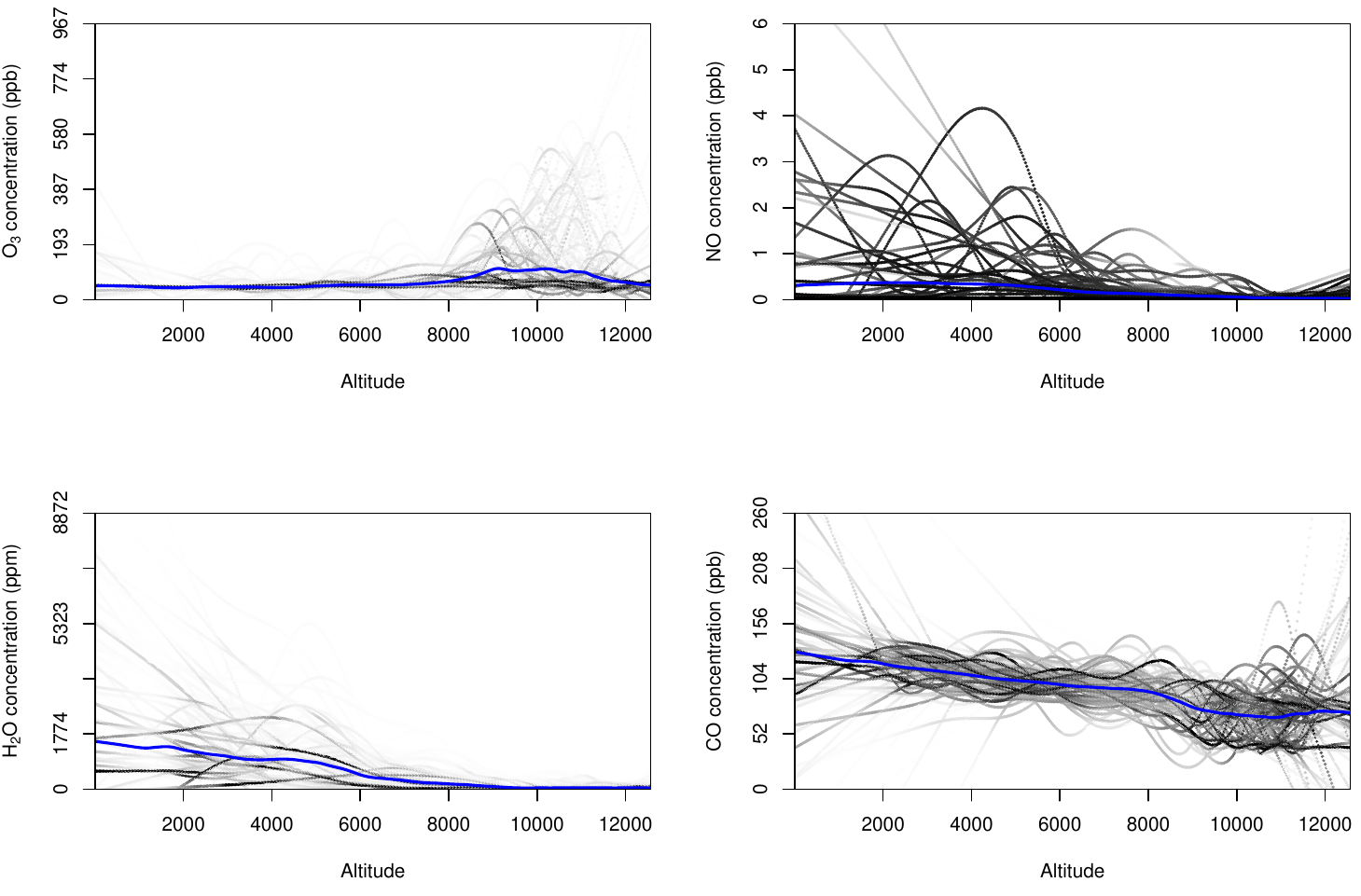}
  \\
  \vspace{1cm}
  \includegraphics[scale=0.7]{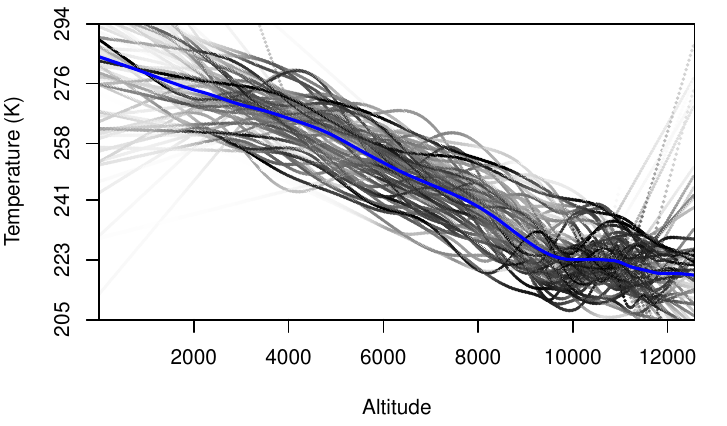}
  \caption{Smooth functional observations $\bm{y}_n$ and $\bm{x}_n$ for $n= 1, \dots, N$. 
  For each altitude value, the points are colored with a grey scale proportional to the wheights $w_{j \, n}(s)$ for $n=1, \dots, N$.
  In blue, the mean function $\bm\mu_j$.
     $O_3$, $NO$, and $CO$ are measured in parts per billion (ppb), while $H_2O$ is measured in parts per million (ppm). Temperature is measured in Kelvin (K).}
  \label{fig:4 mean functions}
\end{figure}

For each variable and each unit, we estimate the underlying true functional form via a non-linear regression on the spatial dimension. 
We use penalized splines to model the relationship between variables and altitude. The level of smoothing is automatically 
selected by the \textbf{gam} function implemented in \textbf{mgcv} library in the statistical programming language \texttt{R}.
Figure \eqref{fig:4 mean functions} shows the resulting smooth functional observations without the noise.

Knowing the full functions $\bm z_n =\{ \bm{y}_n, \bm{x}_n\}$ over the space $\mathcal{S}$, 
and assuming the partial separability of $(\mathcal{C}\bm z)$, it is possible to define a unique system of basis for  $\bm Y$ and $\bm X$, therefore $\varphi_l = \psi_l$ for all $l$. 
From Theorem 2 in \cite{ZapataEtAl_2022}, we know that under partial separability assumption, the eigenvalues of the trace-class 
covariance operator $\mathcal{H}=(p+q)^{-1}\sum^{p+q}_{i=1}\mathcal{C}^{\bm X, \bm Y \mid \bm X}_{ii}$, in the eigenspace spanned by 
$\{\varphi_l\}_{l=1}^\infty$, is equal to the sum of the variances in $\{\langle \bm z_j, \varphi_l\rangle\}_{j=1}^{p+q}$. The eigenbasis of $\mathcal{H}$ are optimal for preserving the maximum amount of overall variability 
between the orthonormal basis system and the functions and for choosing the number of expansion terms which capture a percentage of the total variability of the processes.
The eigenvalues of $\mathcal{H}$ follow a non-increasing order, and for the $L$-th element of the expansion, the corresponding eigenvalue approaches zero.
We evaluate the operator $\mathcal{H}$ fixing grid of 420 points over the altitude, from $1$ meter to $1300$ meters, with steps of 30,
$$({\mathcal{H}}\textbf{z})(t,t') = \frac{1}{5}\sum^4_{j=1} \frac{\sum^{75}_{n=1}w_{j\,n} (t) [ z_{j\,n}(t) - \mu_{j}(t)][ z_{j\,n}(t') - \mu_j(t')]w_{j\,n}(t')}{\sum^{75}_{n=1} w_{j\,n} (t)w_{j\,n} (t')}$$
where $t =1,\dots,420$, $w_{j\,n} (t)$ is a weight, and $\mu_j$ is the mean function of the $j$th variable. The weights are calculated as the inverse of the variances of the estimated functions, and
$\mu_j (t)= \sum^{75}_{n=1}\frac{w_{j\,n} (t) z_{j\,n}(t) }{w_{j\,n} (t)}$.
The eigen-decomposition of $\mathcal{H}\textbf{z}$ gives a discrete version of the basis system. 
The first $5$ basis functions explain more than the $99\%$ of the total variability. Indeed, each variable for each unit is represented by a vector of $5$ scores equal to
$$\bm\gamma_{j\,n}=[\boldsymbol{\varphi}W_{j\,n}\boldsymbol{\varphi}^\top]^{-1}\boldsymbol{\varphi}W_{j\,n} \textbf y_{j\,n} \text{ for } j=1,\dots,4$$
where $\boldsymbol{\varphi} = [\boldsymbol{\varphi}_1 \mid \cdots \mid \boldsymbol{\varphi}_{5}]^\top$ is a $5\times 420$ matrix that collects the values of the eigen-functions, $\textbf y_{j\,n} = \textbf y_{j\,n} -\boldsymbol{\mu}_j$ is the vector of centred observations of length 420, 
and $W_{j\,n}$ is a $420\times 420$ matrix where the diagonal entries are equal to $w_{j\,n}(t)\text{ for } t= 1, \dots, 420$, and the off-diagonal elements are zero. The same is done to estimate the explanatory scores,
$$\bm\chi_{n}=[\boldsymbol{\varphi}W_{X_n}\boldsymbol{\varphi}^\top]^{-1}\boldsymbol{\varphi}W_{X_n} \textbf x_{n},$$ where similarly $W_{X_n}$ is the diagonal matrix of weights, and $\textbf x_{n}$ is the vector of 420 centred temperature values.

\begin{figure}[!ht]
  \centering
  \includegraphics[scale = 0.9]{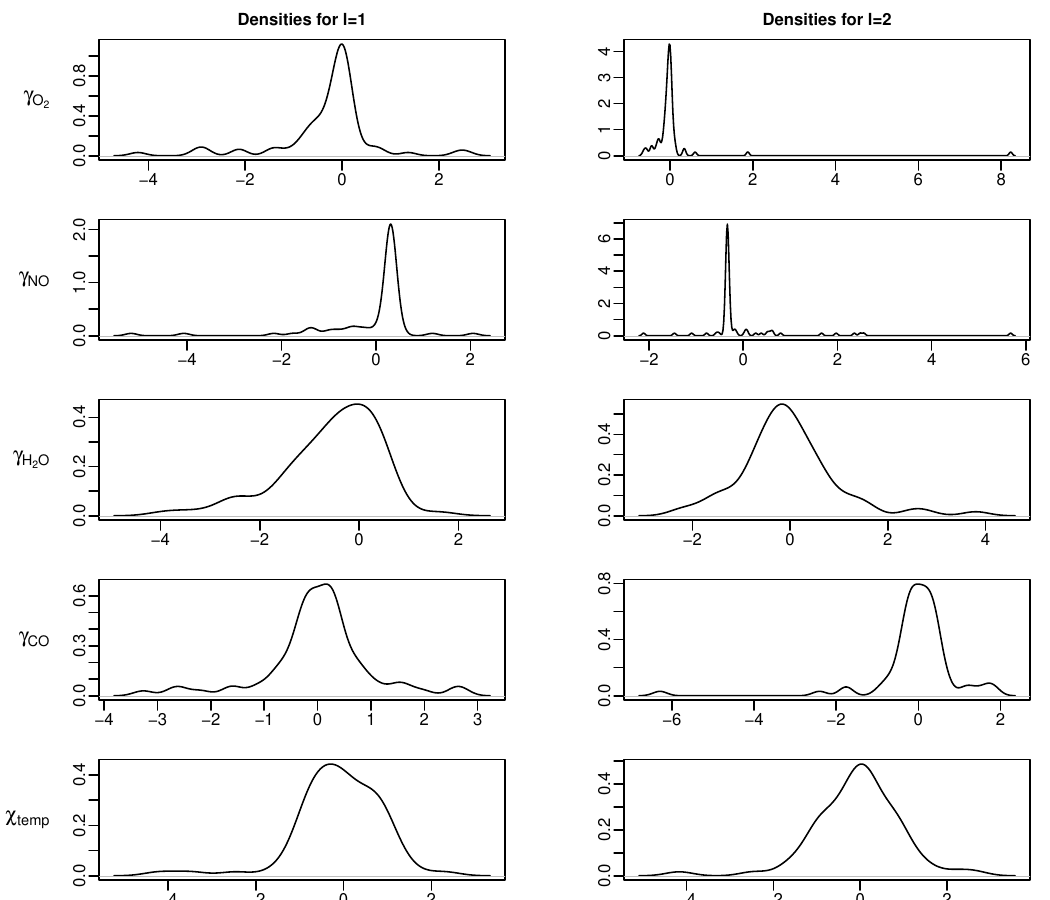}
  \caption{Estimated scores densities for the first and the second term of the expansion.}
  \label{fig:densities}
\end{figure}

Figure \ref{fig:densities} shows the kernel estimation of the densities of the scores for $l=1,2$. 
For each variable and $l$, the scores are divided by their standard deviation. The assumption of Gaussian distributions appears not completely inappropriate, 
although the estimated densities present some asymmetries and kurtosis.

\subsection{fGGRM estimator~\label{sec:estimator_graph-stim}}
The scores $\bm\gamma_{l\,n}$ and $\bm\chi_{l\,n}$ are assumed to follow multivariate Gaussian distributions.
Thanks to the assumption of partial separability of $(\mathcal{C}\bm z)$, the joint distribution is specified as,
\begin{equation}\label{N_q+p}
 \begin{bmatrix} \bm\chi_l \\
                \bm\gamma_l
 \end{bmatrix} 
 \sim
\mathcal N_{q+p}\bigg( 
\bm 0 , 
\begin{bmatrix} \bm \Sigma^{\bm\chi}_l & \bm \Sigma^{\bm\chi}_l  \bm B_l \\ \bm B_l^\top \bm \Sigma^{\bm\chi}_l & \bm B_l^\top\bm \Sigma^{\bm\chi}_l \bm B_l+ \bm \Sigma^{\bm\epsilon}_l\end{bmatrix} \bigg) \, .
\end{equation}
The precision matrix of that density is given as,
\begin{equation}\label{prec.Z}
     \Theta_l = \begin{bmatrix} \bm\Theta^{\bm\chi}_l + \bm B_l \bm\Theta^{\bm\gamma}_l \bm B_l^\top & -\bm B_l \bm\Theta^{\bm\gamma}_l \\ -\bm\Theta^{\bm\gamma}_l \bm B_l^\top & \bm\Theta^{\bm\gamma}_l \end{bmatrix} \, ,
\end{equation}
where $\bm{\Theta}^{\bm{\chi}}_l$ and $\bm{\Theta}^{\bm{\gamma}}_l$ are the inverses of the marginal covariance matrices of $\bm{\chi}_l$ and $\bm{\gamma}_l$, respectively. Because of partial separability, the densities \eqref{N_q+p} are independent over $l$. 

We propose to estimate a functional Gaussian graphical regression model from $\{\bm\chi_{h\,n}, \bm\gamma_{l\,n}\}_{l\,h\,n}$, using the following double-penalized estimator, named \textit{functional joint conditional graphical lasso} estimator,
\begin{equation}\label{eqn:fjcglasso}
\{\pmb{\widehat B}\}, \{\widehat\Theta^{\bm\gamma}\} = \underset{\{\pmb{ B}\}, \{\Theta^{\bm\gamma}\}}{\operatorname{argmax}}\sum_{l = 1}^L \left\{\log\det ( \Theta^{\bm\gamma}_l) - \tr{( \bm S(\pmb B_l)\Theta^{\bm\gamma}_l)}\right\} - \nu P_1(\{\pmb B\}) - \rho P_2(\{\Theta^{\bm\gamma}\})\, ,
\end{equation}
where 
\begin{equation}
\label{S.B}
\bm S(\bm B_l) = \frac{1}{N} \sum^N_{n=1} (\bm\gamma_{l\,n}- \bm B_{l}\bm\chi_{l\,n} )^\top (\bm\gamma_{l\,n}- \bm B_{l}\bm\chi_{l\,n} ) \, .
\end{equation}
The penalty functions $P_1(\cdot)$ and $P_2(\cdot)$ are convex functions that encourage sparsity in each expansion matrix and specific forms of similarity across the expansion elements.
We use the group lasso penalty functions. The penalty for the regression coefficient matrices and for the precision matrices are
$$ P_1(\{\pmb B\}) = \sum_{i = 1}^p\sum_{k = 1}^q \left(\sum_{l=1}^{L} b^2_{l\,ki}\right)^{1/2},\hspace{0.2cm}\textrm{ and }  \hspace{0.2cm} P_2(\{\Theta\}) = \sum_{i\ne j}^p \left( \sum_{l=1}^{L}\theta_{l\,ij}^2 \right)^{1/2} . $$ 
Thus, the desired edge sets can be estimated by
$$ \widehat\mE^{\bm X \bm Y}_{\nu\rho}  = \left\{(k, i):\sum_{l = 1}^L  (\hat b^{\rho \nu}_{l\,ki})^2 > 0\right\}\hspace{0.2cm}\textrm{ and }  \hspace{0.2cm} \widehat{\mE}^{\bm Y}_{\nu\rho} = \left\{(i, j):\sum_{l = 1}^L( \hat\theta^{\rho\nu}_{l\,ij})^2 > 0\right\} \, ,
$$
where $\hat b^{\rho \nu}_{l\,ki}$ and $\hat\theta^{\rho\nu}_{l\,ij}$ are the matrices entries estimated under the couple of penalization values $(\rho,\nu)$.

\section{Goodness of Fit and Model Selection~\label{sec:goodness_of_fit} }
In this section, we turn our attention to the task of assessing the performance and selecting optimal configurations for the proposed model.
The estimation is performed using the double-penalized optimization problem in \eqref{eqn:fjcglasso}. The penalty parameters are individually tuned 
and jointly influence the estimation of both the precision matrices for the response variables and the regression matrices.
The central challenge in conditional independence graph estimation is to determine which precision and regression matrices entries are to be non-zero. In \textit{lasso} estimation, the tuning parameters control the sparsity of the network, and the problem of selecting the tuning parameters values coincides with model selection.

In the remaining part of this section, we first explain how we estimate the Kullback-Leibler divergence for the fcGGM estimator. Then, we discuss the tools for selecting tuning parameters, introducing the joint version of KLCV.

\subsection{Goodness of fit}
Kullback Leibler (KL) divergence serves as a non-symmetric measure to quantify the discrepancy between two probability distributions \citep{Penny_2001,JoramEtAl_2016}. 
It gauges the information loss that occurs when one probability distribution is employed to approximate another.
The KL measure is defined as:
\begin{equation}\label{KL}
KL(f_1 \mid f_2)= \Ev_{f_1} [ \log \left(f_1 \right)- \log \left( f_2\right) ]
\end{equation}
where $f_1$ and $f_2$ are two probability density functions. 

In our study, we aim to evaluate the information loss between the true multivariate normal distribution \eqref{N_q+p} and the estimated distributions derived from \eqref{eqn:fjcglasso} under various combinations of tuning parameters values.
The couple $(\rho, \nu)$ dictate the strength of penalization, leading to distinct estimations of $\{\widehat\Theta^{\bm\gamma}\}$, $\{\widehat\Theta^{\bm\chi}\}$, and $\{\widehat{\bm B}\}$, which together specify the normal distribution we seek to compare with the true distribution.
As shown in \cite{Penny_2001}, the KL divergence between two zero-mean multivariate normal distributions can be written as
\begin{equation}\label{KLmultivariatenormal}
KL_{\mathcal N_m}[\bm\Theta \mid \widehat{\bm\Theta}] = \frac{1}{2}\{ \tr(\bm\Theta^{-1}\widehat{\bm\Theta}) - \log\mid\bm\Theta^{-1}\widehat{\bm\Theta}\mid-m\}.
\end{equation}
As shown in Section \eqref{sec:fcggm}, the density function of the joint vector can be factorized as the product between the explanatory and the conditioned response density functions
\begin{equation}\label{factorized.density}
f(\bm\chi_l, \bm\gamma_l \mid \bm\chi_l; \bm\Theta_l) = f(\bm\chi_l; \bm\Theta^{\bm\chi}_l,  \bm B_l) f(\bm\gamma_l \mid \bm\chi_l; \bm\Theta^{\bm\gamma}_l, \bm B_l) \, .
\end{equation}
The KL loss can be calculated as the sum of $L$ divergences
\begin{align}\label{KL_Z}
&KL_{\mathcal N_{q+p}}\left[\{\bm\Theta^{\bm\chi} \}\{\bm\Theta^{\bm\gamma} \}\{\bm B \}\mid
\{\widehat{\bm\Theta}^{\bm\chi}\}_{\rho\,\nu}\{\widehat{\bm\Theta}^{\bm\gamma}\}_{\rho\,\nu}\{\widehat{\bm B}\}_{\rho\,\nu}\right] =
\\&= \frac{1}{2}\sum^L_{l=1}\bigg\{ \tr\left(\bm\Psi_l^{-1}\widehat{\bm\Psi}_{l\,\rho\nu}\right) - \log\mid\bm\Psi_l^{-1}\widehat{\bm\Psi}_{l\,\rho\nu}\mid + 
\notag\\& + \tr\left(\bm\Theta^{\bm\gamma \, -1}_l\widehat{\bm\Theta}^{\bm\gamma}_{l\,\rho\nu}\right) - 
\log\mid\bm\Theta^{\bm\gamma \, -1}_l\widehat{\bm\Theta}^{\bm\gamma}_{l\,\rho\nu}\mid 
-\left(q+p\right) \bigg\} \notag \, ,
\end{align}
where $\bm\Psi_l$ is the left-top block in \eqref{prec.Z}.

\subsection{Model selection methods}\label{sec:model_selection_methods}
The most common approach for comparing and selecting statistical models involves statistics based on the concept of expected KL divergence between the model under examination and the true model. When the true model is elusive, the common strategy is to estimate this discrepancy by considering the sum of two distinct terms. One term is the maximized log-likelihood of the model being tested. This term serves as an indicator of how well the model fits the observed data. The second term, on the other hand,  serves a contrasting purpose. It is designed to mitigate the risk of overfitting, which can occur when a model is too complex and overly tuned to the observed data. 
This term offers an estimate of the bias that emerges when the true distribution is approximate by the estimated parameters.
In essence, this two-term approach strikes a balance between model fit and the risk of overfitting, favouriting a more robust and well-informed model selection process.

Assuming to have $N$ realization of the processes $\bm Y$ and $\bm X$, the log-likelihood of density \eqref{factorized.density} is
\begin{align*}
  & \ell\left(\bm\chi, \bm\gamma\mid\bm\chi; \left\{\widehat{\bm\Psi}\right\}, \left\{\widehat{\bm\Theta}^{\bm\gamma}\right\} \right)  =  \frac{N}{2} \sum^L_{l=1} \ell_l\left( \bm\chi; \bm\hat\Psi_l \right) + \ell_l\left( \bm\gamma; \widehat{\bm\Theta}^{\bm\gamma}_l \right) \\
   & =  \frac{N}{2} \sum^L_{l=1} \log \mid\widehat{\bm\Psi}_l\mid -\tr\left(\bm S^{\bm\chi}_l \widehat{\bm\Psi}_l\right) + \log\mid\widehat{\bm\Theta}^{\bm\gamma}_l\mid -\tr\left(\bm S^{\bm\gamma}_{\widehat{\bm B}_l} \widehat{\bm\Theta}^{\bm\gamma}_l\right) \, ,
\end{align*}
where $ \bm S^{\bm\chi}_l = N^{-1} \sum^N_{n=1} \bm\chi_{l\,n}\bm\chi^\top_{l\,n}$ and
$\bm S^{\bm\gamma}_{\widehat{\bm B}_l} = N^{-1}\sum^N_{n=1}( \bm\gamma_{l\,n} - \bm\chi_{l\,n} \widehat{\bm B}_{l}) ( \bm\gamma_{l\,n} - \bm\chi_{l\,n} \widehat{\bm B}_{l} )^\top.$
The way to estimate the bias term fundamentally characterizes the model selection statistic. Both AIC and BIC penalize the log-likelihood by the degrees of freedom scaled by a parameter. For graphical models, these criteria have the equations
$$ AIC(\widehat{\mathcal E}_{\rho\nu}) = -2\ell\left(  \bm\chi, \bm\gamma\mid\bm\chi; \left\{\widehat{\bm\Psi}\right\}_{\rho\nu}, \left\{\widehat{\bm\Theta}^{\bm\gamma}\right\}_{\rho\nu}\right)+ 2 \mid \widehat{\mathcal E}_{\rho\nu}\mid $$
and $$ BIC(\widehat{\mathcal E}_{\rho\nu}) = -2 \ell\left(  \bm\chi, \bm\gamma\mid\bm\chi; \left\{\widehat{\bm\Psi}\right\}_{\rho\nu}, \left\{\widehat{\bm\Theta}^{\bm\gamma}\right\}_{\rho\nu}\right) + \log (N)\mid \widehat{\mathcal E}_{\rho\nu}\mid$$
where $\mid \widehat{\mathcal E}_{\rho\nu}\mid$ is the number of links in the graph.
While BIC is known to be consistent for a fixed number of parameters and increasing sample size, it may not necessarily select a parsimonious model when the model space is large.
The extended BIC (eBIC), introduced in \cite{Chen&Chen}, considers both the number of non-zero estimated parameters and the complexity of the model space. eBIC is particularly useful for variable selection in problems with moderate sample sizes and a large number of variables. It differs from BIC and AIC by an additional penalty term that controls the prior probability of sparse models
$$eBIC(\widehat{\mathcal E}_{\rho\nu})= - 2 \ell  \left(   \bm\chi, \bm\gamma\mid\bm\chi; \left\{\widehat{\bm\Psi}\right\}_{\rho\nu}, \left\{\widehat{\bm\Theta}^{\bm\gamma}\right\}_{\rho\nu} \right) +  log(N) \mid \widehat{\mathcal E}_{\rho\nu} \mid +  4 g \mid \widehat{\mathcal E}_{\rho\nu} \mid log(p+q) \, ,$$
where $p+q$ is the total number of nodes in the graph. The value of $g$ is manually set and larger values of it result in sparser models. When $g = 0$, eBIC is equivalent to the ordinary BIC. The eBIC remains consistent even when the number of parameters grows to infinity with the sample size  \citep{Chen&Chen,WysockiAndRhemtulla}.
This criterion effectively controls the false discovery rate and often shows good performances in terms of graph recovery.

When the goal is to obtain a model with good predicting power, cross-validation is the gold standard.
The KLCV leverages the cross-validation of the log-likelihood loss to estimate the KL divergence. This criterion estimates the bias term by approximating leave-one-out-cross validation and offers a computationally fast alternative to cross-validation.
This technique often outperforms other methods, particularly when the sample size is small.
The proposed joint KLCV is defined as:
\begin{align*}
 jKLCV\left(\left\{\widehat{\bm\Psi}\right\}_{\rho\nu}, \left\{\widehat{\bm\Theta}^{\bm\gamma}\right\}_{\rho\nu} \right) & = - \frac{1}{2L} \ell\left( \bm\chi, \bm\gamma\mid\bm\chi; \left\{\widehat{\bm\Psi}\right\}_{\rho\nu}, \left\{\widehat{\bm\Theta}^{\bm\gamma}\right\}_{\rho\nu} \right) +\\
&+ \frac{1}{NL(NL-1)} \sum^L_{l=1} \sum^N_{n=1}\widehat{\mbox{bias}}^{\bm\chi_{n}}_{l\,\rho\nu} + \widehat{\mbox{bias}}^{\bm\gamma_{n} \mid \bm\chi}_{l\,\rho\nu}.
\end{align*}
Adapting the results in \cite{WitEtAl_15}, for each realization of the process, the bias terms are estimated as,
$$\widehat{\textrm{bias}}^{\bm\chi_{n}}_{l\,\rho\nu} = \text{vec}\left[ \left( \widehat{\bm\Psi}_{l\,\rho\nu}
^{-1} - \bm S_{l\,\rho\nu}^{\bm\chi_{n}}  \right) \odot \mathtt I^{\bm\chi}_{l\, \rho\nu} \right]^\top \text{vec}\left[ \widehat{\bm\Psi}_{l\, \rho\nu} \left( \left( \bm S_{l\,\rho\nu}^{\bm\chi} - \bm S_{l\,\rho\nu}^{\bm\chi_n}  \right) \odot \mathtt I^{\bm\chi}_{l\,\rho\nu} \right) \widehat{\bm\Psi}_{l\,\rho\nu} \right]$$
and
$$\widehat{\textrm{bias}}^{\bm\gamma_{n}\mid\bm\chi}_{l\,\rho\nu} = \text{vec}\left[ \left({\widehat{\bm\Theta}^{\bm\gamma \, -1}}_{l\,\rho\nu}- \bm S^{\bm\gamma_n}_{\widehat{\bm B}_{l\,\rho\nu}} \right) \odot \mathtt I^{\bm\gamma}_{l\,\rho\nu} \right]^\top \text{vec}\left[ \widehat{\bm\Theta}^{\bm\gamma}_{l\,\rho\nu} \left( \left( \bm S^{\bm\gamma}_{\widehat{\bm B}_{l\,\rho\nu}} - \bm S^{\bm\gamma_n}_{\widehat{\bm B}_{l\,\rho\nu}} \right) \odot \mathtt I^{\bm\gamma}_{l\,\rho\nu} \right) \widehat{\bm\Theta}^{\bm\gamma}_{l\,\rho\nu} \right]$$
where $\mathtt I^{\bm\chi}_{l\,\rho\nu}$ and $\mathtt I^{\bm\chi}_{l\,\rho\nu}$ are the indicator matrices, whose entry is $1$ if the corresponding entry in $\widehat{\bm\Psi}_{l\,\rho\nu} $ or $\widehat{\bm\Theta}^{\bm\gamma}_{l\,\rho\nu} $ is non-zero, and $0$ if the corresponding entry is zero. 
The observed covariance matrices for the unit $n$ are $\bm S^{\bm\chi_n}_{l} = \bm\chi_{l\,n} \bm\chi_{l\,n}^\top$ and $\bm S^{\bm\gamma_n}_{\widehat{\bm B}_{l\,\rho\nu}}= (\bm\gamma_{l\,n} - \bm\chi_{l\,n} \widehat{\bm B}_{l\,\rho\nu} )  (\bm\gamma_{l\,n} - \bm\chi_{l\,n} \widehat{\bm B}_{l\,\rho\nu} )^\top$.

In Section \ref{sec:simulation}, the AIC, eBIC and jKLCV are compared. In general, if the aim is graph identification, then the eBIC is appropriate. On the other hand, if the aim is to find a model with good predictive power, jKLCV and AIC are preferred.

\section{Simulation study~\label{sec:simulation}}

In this section, we study the finite-sample performance both of the propose fGGRM estimator and the jKLCV measure of goodness-of-fit.

\subsection{Effect of sample size on predictive and graph recovery power}

In this study, the following setting was used to simulate independent samples of random responses and covariance functions.

\begin{figure}[t!]
\centering
\makebox{
\includegraphics[scale=.5]{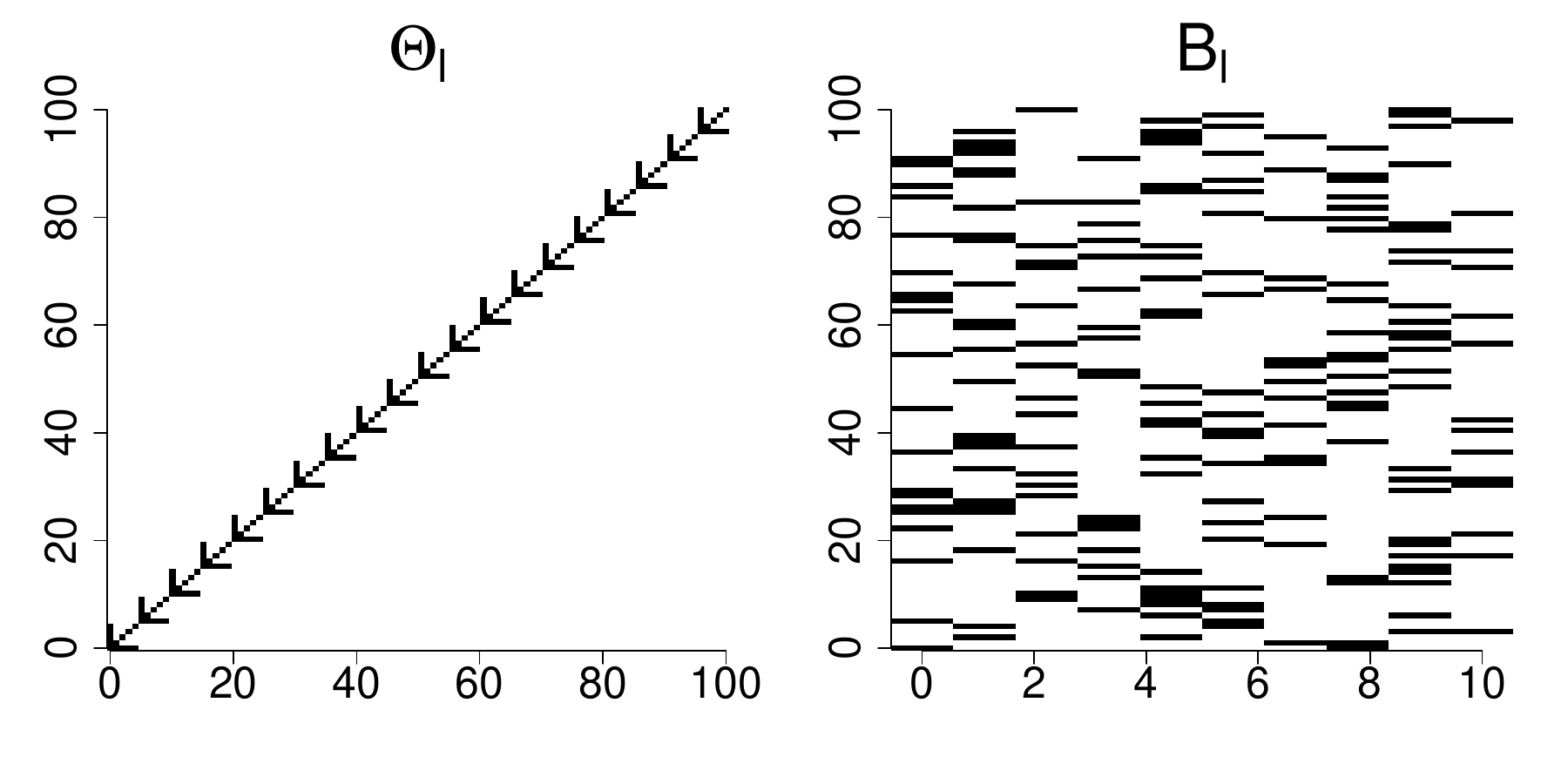}
}
\caption{Example of sparse structure for $\Theta_l$ and $B_l$ for $p=100$, $q=10$ and for each $l=1, \ldots, L$.
\label{fig:structure}}
\end{figure}

We consider a model with $p = 100$ functional responses and $q = 10$ functional covariates which, under the assumption of partial separability of the covariance operator, can be represented using the multivariate Karhunen-Loève expansion~(\ref{exp.YX}). In our study, we set $L = 3$ and, according to model~(\ref{N_q+p}), for each $l$, the vector of random scores $(\bm\chi^\top_l, \bm\gamma_l^\top)^\top$ was drawn from a multivariate gaussian distribution with zero expected value and precision matrix specified as in~(\ref{prec.Z}), where $\Theta^{\bm\chi}_l$ was let equal to the identity matrix whereas $\Theta^{\bm\gamma}_l$ was constructed using the following star configuration: $\theta^{\bm\gamma}_{l\;r(r+s)}$ is non-zero only when $r = 1, 6, 11, \ldots$ and $s = 1,\ldots, 4$ (see Figure~\ref{fig:structure}). The values of the non-zero entries $\theta^{\bm\gamma}_{l\;r(r+s)}$ were sampled from a uniform distribution $U([-0.5, -0.4]\cup [+0.4, +0.5])$. Similarly to~\cite{ZapataEtAl_2022}, each covariance matrix $\Sigma_l$ was scaled by the factor $1/l^{0.2}$ to guarantee monotonically decreasing traces. The matrices of regression coefficients $\bm B_l$ were generated in such a way that only two randomly selected functional covariates affect each functional response. More specifically, the values of the non-zero regression coefficients were drawn from $U([-1.4, -1.0]\cup [+1.0, +1.4])$. The previous setting was used to draw samples of $N \in\{50, 100\}$ random response and covariate functions.

We compare the behavior of the proposed fGGRM model with two alternative competitors. The first, is the estimator proposed in~\cite{ZapataEtAl_2022} which, in our study, consists in fitting the jglasso model, with group lasso penalty function, using only the scores $\bm\gamma_l$. This model was used to study how the lack of considering the functional predictors affects the estimates of the precision matrices. The second competitor, which we refer as ``naive'' estimator, is defined as follows: first, for each $l$, we fit a conditional glasso model using $\bm\gamma_l$ and $\bm\chi_l$ then, the edge-sets $\mathcal E^{\bm Y}$ and $\mathcal E^{\bm X\bm Y}$ are estimated using the OR rule, i.e., $(i,j)\in\widehat{\mathcal E}^{\bm Y}$ if and only if exists a $\hat\theta^\gamma_{l\;ij}\neq 0$, with $l = 1,\ldots, L$ and, similarly, $(k,i) \in \widehat{\mathcal E}^{\bm X\bm Y}$ if and only if exists at least an index $l$ such that $\hat b_{l\;ki}\neq0$. 

The considered models were compared both in terms of ability of recovering the edge-sets $\mathcal E^{\bm Y}$ and $\mathcal E^{\bm X\bm Y}$, and the accuracy in estimating the parameters of the true model generating data. In particular, in each simulation run, to evaluate $\{\widehat\Theta^{\bm\gamma}\}$ and the corresponding $\widehat{\mathcal E}^{\bm Y}$, taking into account the effects coming from $\{\bm{\widehat{B}}\}$, we used the following strategy. For each method, we computed a decreasing sequence of ten $\nu$-values and, for each of these values, a decreasing sequence of ten $\rho$-values was used to fit the models. The largest $\nu$ and $\rho$ values were computed as specified in~\cite{SottileEtAl_2022}. The resulting sequences of estimates of the precision matrices were summarized using the area under the ROC curve (AUC) and the following estimate of the average mean-square error:
\[
\widehat{aMSE} = \frac{1}{L}\sum_{l = 1}^L\|\widehat\Theta^{\bm\gamma}_l - \Theta^{\bm\gamma}_l\|^2_F.
\]
The previous strategy was reversed to evaluate $\{\bm{\widehat{B}}\}$ taking into account $\{\widehat\Theta^{\bm\gamma}\}$, i.e., first a decreasing sequence of ten $\rho$-values was computed and, for each of these values, the models were fitted using a decreasing sequence of $\nu$-values. As before, the behavior of the resulting estimators was evaluated in terms of AUC and estimate of the average mean-square error. 

In the remaining part of this section we investigate the results coming from 100 simulation runs.

\begin{figure}[tbp!]
\centering
\makebox{
\includegraphics[scale=.36]{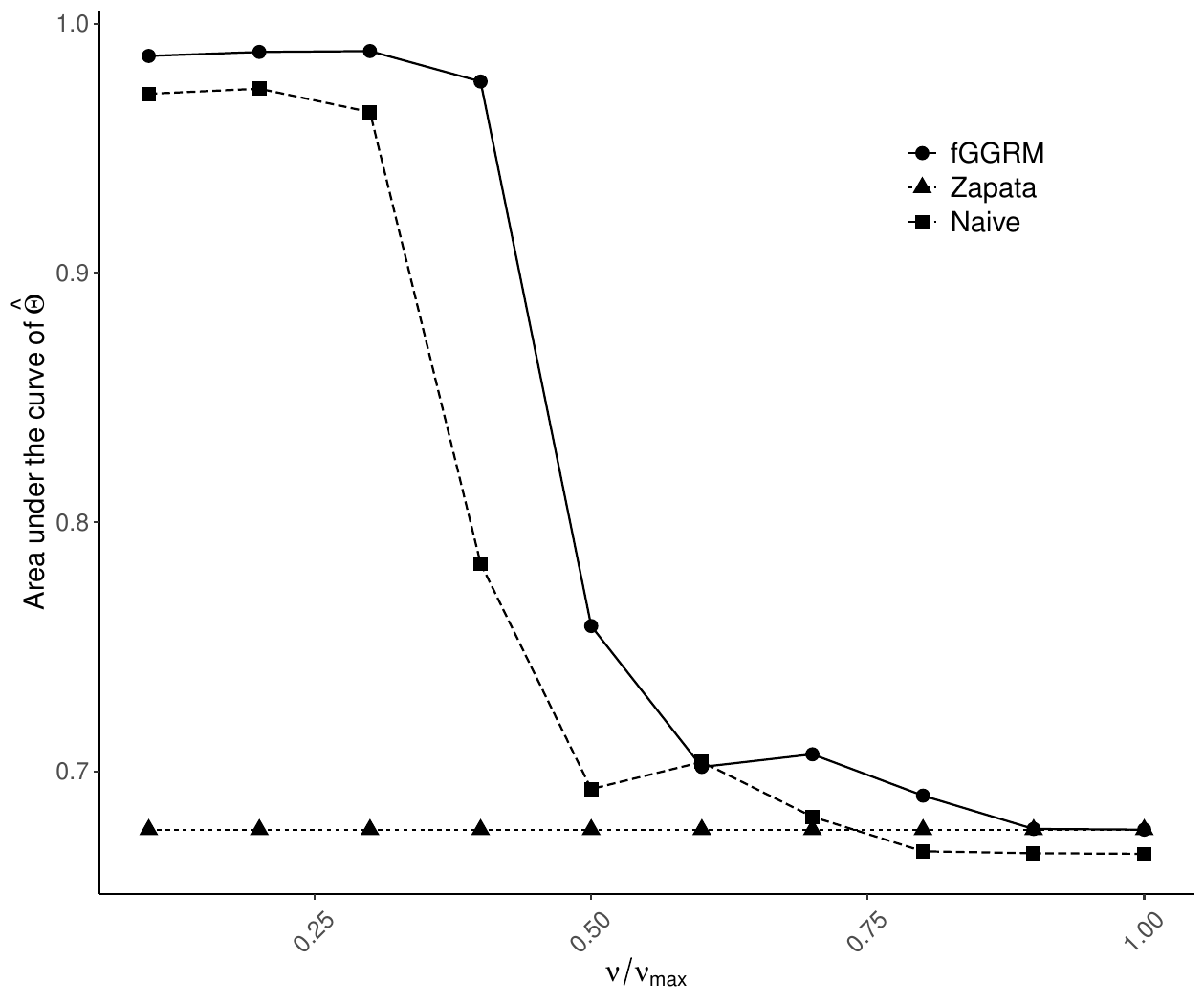}
\includegraphics[scale=.36]{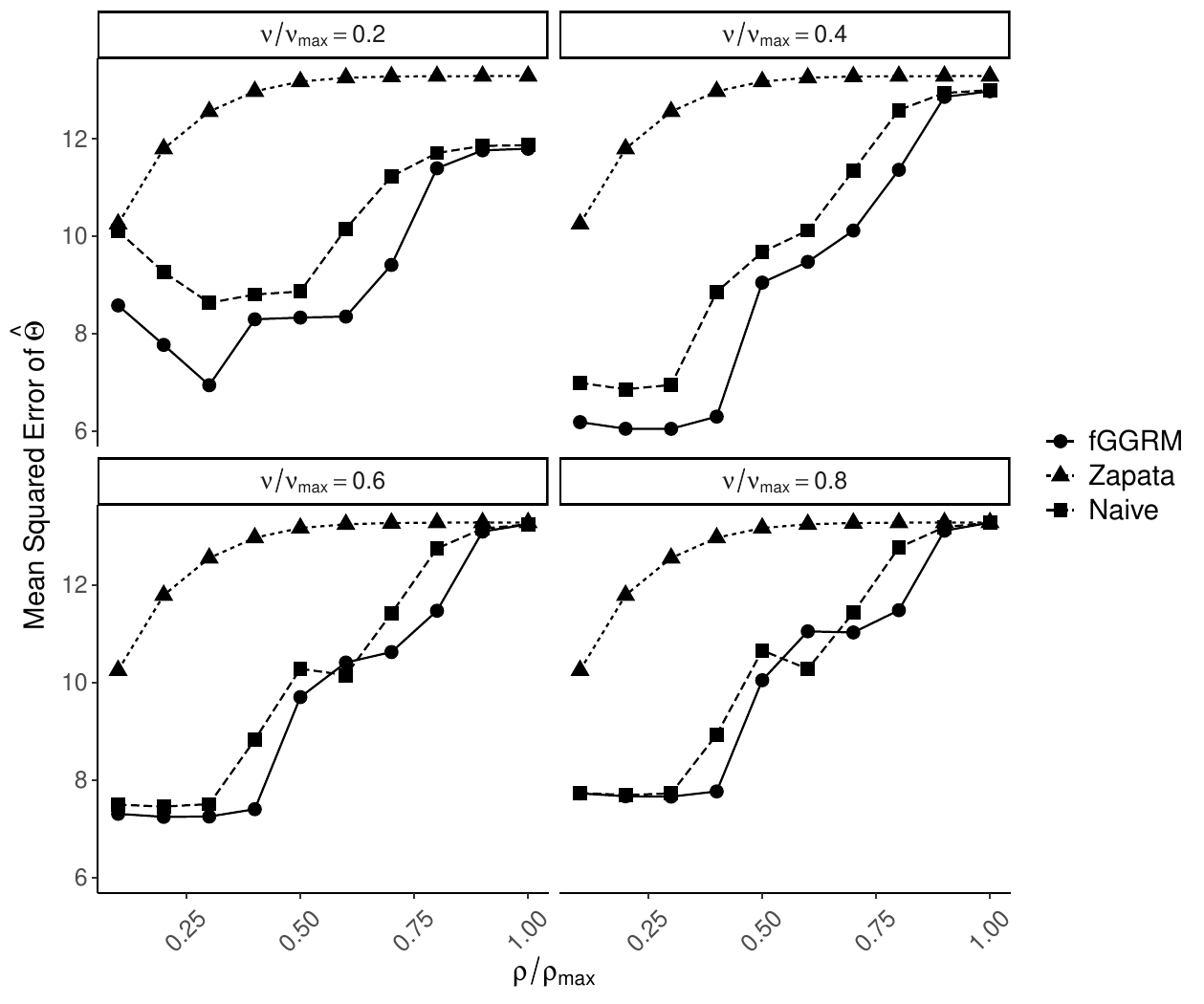}
}
\makebox{
\includegraphics[scale=.36]{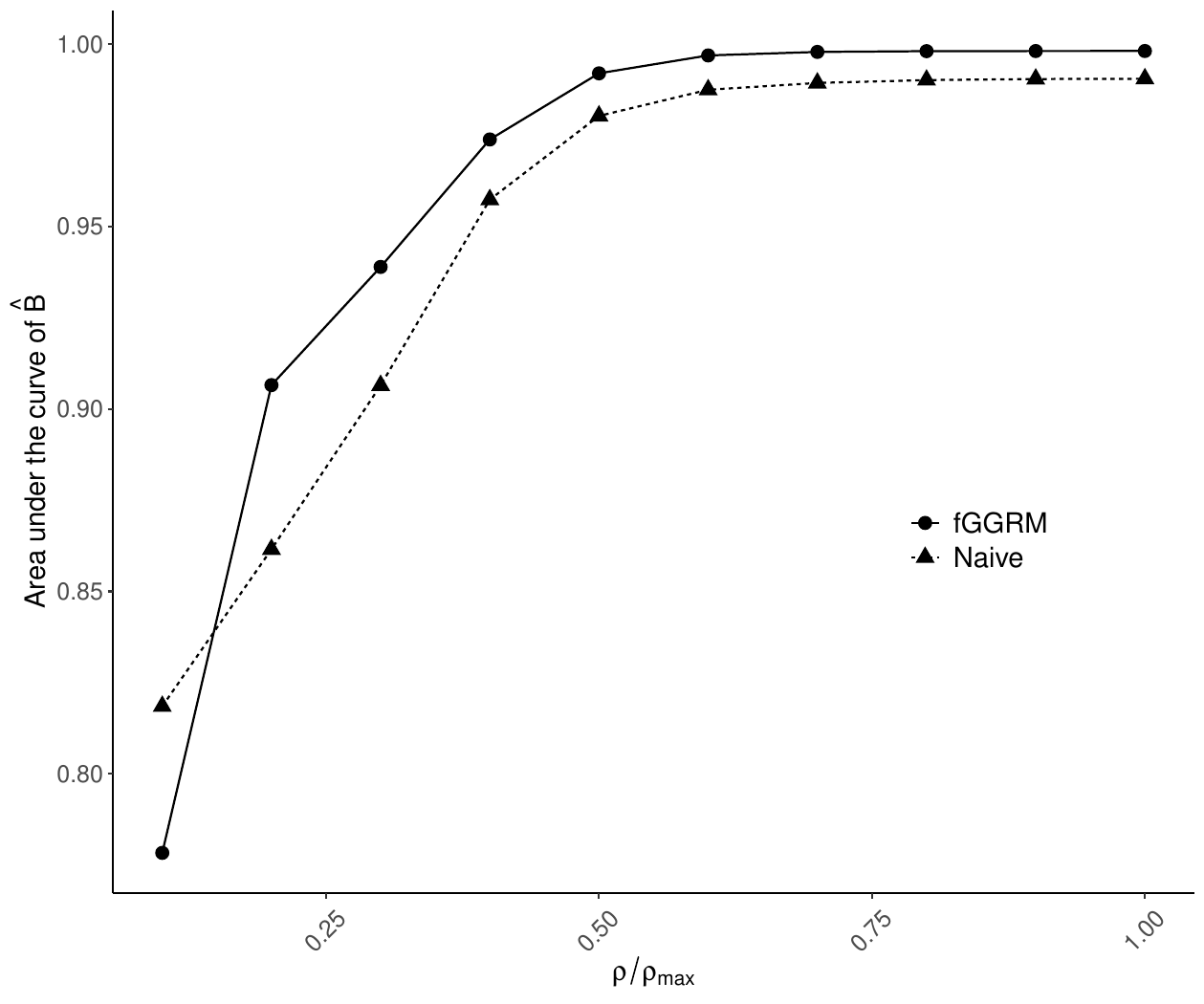}
\includegraphics[scale=.36]{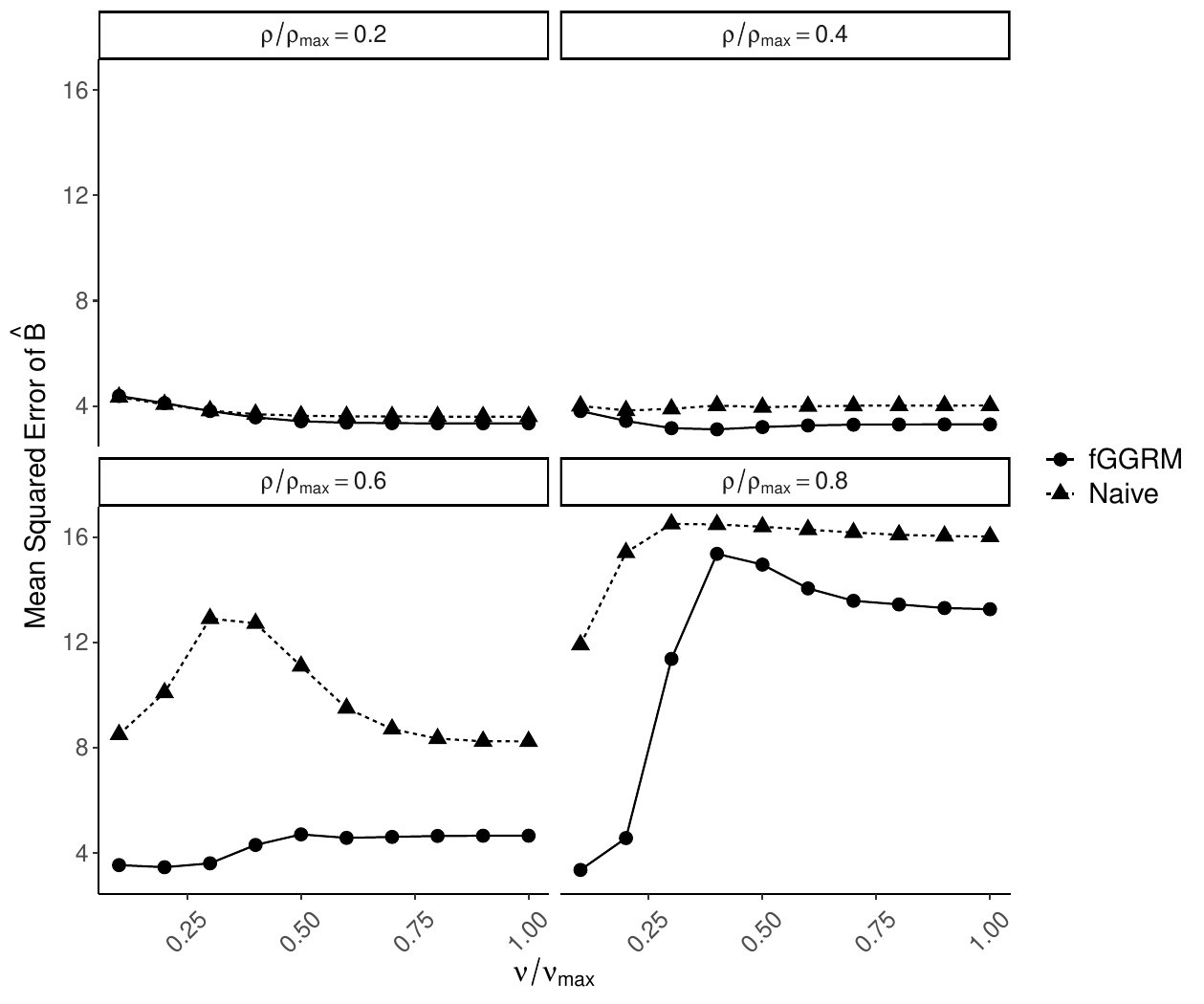}
}
\caption{Simulation study with $N=50$, $p = 100$ and $q = 10$. The three lines in the upper panels correspond to \texttt{fGGRM}, the model proposed in~\cite{ZapataEtAl_2022} and the ``naive'' estimator. In the bottom panels the method proposed in~\cite{ZapataEtAl_2022} was removed because it not consider the functional predictors. Left panels refer to the AUC curves for the edge-sets $\mathcal E^{\bm Y}$ and $\mathcal E^{\bm X\bm Y}$, respectively. Right panels refer to the MSE curves for $\widehat{\Theta_l}$ and $\widehat{B_l}$. The AUC curves are evaluated over the ratio $\nu/\nu_\text{max}$ and $\rho/\rho_\text{max}$, respectively. The MSE curves for $\widehat{\Theta_l}$ are computed over the the ratio of ten $\rho$ -values keeping fixed the structure of the regression coefficient matrix by some ratio $\nu/\nu_\text{max}$. The MSE curves for $\widehat{B_l}$ are computed over the the ratio of ten $\nu$-values keeping fixed the structure of the precision matrix by some ratio $\rho/\rho_\text{max}$.\label{fig:auc_mse_n50}}
\end{figure}

The results of the simulation study with $N=50$ are shown in Figure~\ref{fig:auc_mse_n50}. Looking at the AUC curves, the proposed method consistently outperforms the competitor. Similar results are shown when looking at the $\widehat{aMSE}$ curves, where our proposal always seems to outperform the other methods. These results suggest how including functional responses and predictors in a single model and applying a double penalty could provide more insight in such applications. Similar considerations can be made for $N=100$ (results not shown).
%

\subsection{Comparison of jKLCV with other methods\label{subsec:comparison_KLCV}}

In this section, we evaluate the ability of various model selection methods to select models with low KL loss or with a good graph recovery power.

\paragraph{jKLCV has good predictive power.}
Table~\ref{table.PredictivePower} showcases the performance of different model selection criteria in terms of KL loss with varying sample sizes.
The values in Table~\ref{table.PredictivePower} (left) are obtained by evaluating \eqref{KL_Z} on the estimates selected by the information criteria presented in Section \eqref{sec:model_selection_methods}. When the sample size is small, the $jKLCV$ outperforms the other bias approximations.
When $N$ increases, the three methods present smaller differences, and the AIC slightly outperforms the others. The main reason is that the explicit bias estimation in jKLCV is subject to sampling error, whereas the asymptotic correction term in the AIC is constant.

\begin{table}[t!]
  \caption{\emph{Left}: KL loss using different model selection criteria. \emph{Right}: Recovery accuracy adopting different model selection criteria. The results are the medians over $100$ simulations with $L = 3$ and varying the sample sizes.\label{table.PredictivePower}}
  \begin{minipage}{0.5\textwidth}
  \begin{center}
  \begin{tabular}{rccc}
  N  &  jKLCV&  AIC & eBIC\\\hline
  10  & \textbf{35.55} & 54.78 &   36.20    \\
  100  &  15.61   & \textbf{15.11} & 15.76  \\
  \hline
  \end{tabular}
  \end{center}
  \end{minipage}%
  \begin{minipage}{0.5\textwidth}
      \begin{center}
  \begin{tabular}{rccc}
  N  &  jKLCV&  AIC & eBIC\\\hline
  10 &  0.73 & 0.48 & \textbf{0.75}   \\
  100 & 0.95 &  \textbf{0.96} &  0.94    \\
  \hline
  \end{tabular}
  \end{center}
  \end{minipage}
  
  \end{table}

\paragraph{eBIC has good graph recovery power.}
Table~\ref{table.PredictivePower} (right) shows that with a small sample size, the $eBIC$ has a stronger graph recovery power than $AIC$ and slightly outperforms the $jKLCV$. When $N$ is high, all three methods present a good ability to classify the matrices' entries correctly.

\section{Structural interaction of pollutants in the air}\label{sec:illustration}
\begin{figure}[!h]
  \centering
  \includegraphics[scale = .7]{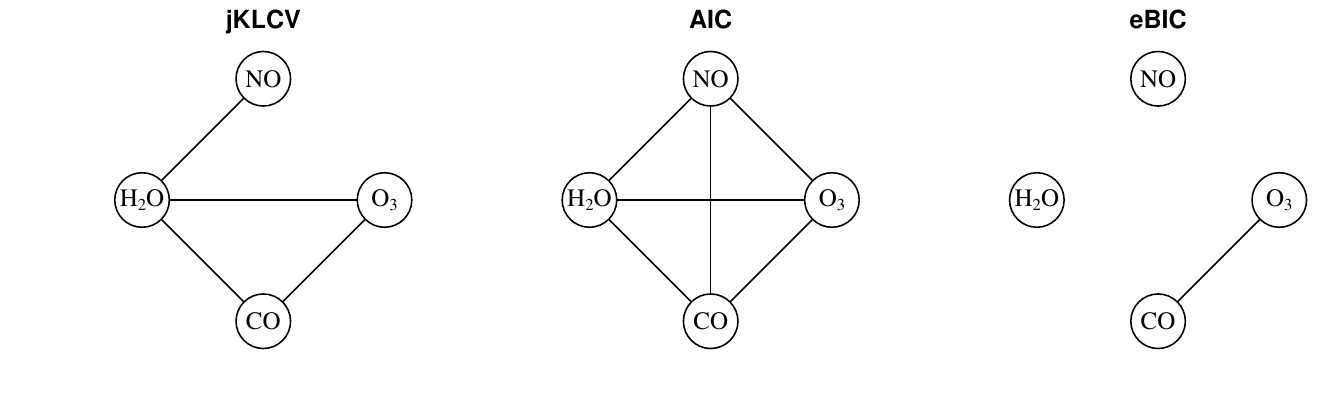}
  \caption{Undirected networks estimated using jKLCV, AIC and eBIC.}
  \label{fig:nets}
\end{figure}

In this section, we present the results of air pollution analysis using the functional Gaussian graphical regression model described in this paper. 
The optimization process is performed iteratively to recover the edge set. In each iteration, one of the two tuning parameters is fixed while the other is varied across $101$ values, ranging from its maximum to zero. 
The total number of iterations is $3$. 
Figure \ref{fig:nets} and Table \eqref{tab:beta} show the graph structures selected using the $jKLCV$, $AIC$ and $eBIC$ criteria at each iteration. 
\begin{table}[!h]
  \centering
  \begin{tabular}{|c|c|c|c|c|}
  \hline
      & $O_3$ & $NO$  & $H_2O$ & $CO$ \\ \hline
  AIC   &   1  &  1   &    1  &  1  \\ \hline
  eBIC  &    1 &   1  &    1  &   1 \\ \hline
  jKLCV &   0  &    0 &   1   &    0\\ \hline
  \end{tabular}
  \caption{Directed links selected using jKLCV, AIC and eBIC.}
  \label{tab:beta}
\end{table}
Adopting AIC leads to selecting a full graph for both directed and undirected sets of links, 
implying that all variables are conditionally dependent on each other given the rest of the variables in the model. 
eBIC selects a full directed network and a very parsimonious undirected network with only one link between $CO$ and $O_3$.
The link between $CO$ and $O_3$ reflects the significant role that $CO$ plays in the formation of $O_3$ in the troposphere, as highlighted in \cite{VoulgarakisEtAl_11} and \cite{ChoiEtAl_17}. 
$CO$ is often used to estimate the amount of Ozone produced by anthropogenic sources. 
However, full graphs or graphs with only one link tend to be less interpretable in practical applications. 
The model selected by jKLCV provides a balanced perspective. 
It suggests a directed link from temperature to $H_2O$, and an undirected network where $H_2O$ maintains all its links, along with the link between $CO$ and $O_3$.
The model selected by jKLCV reflects the critical role of $H_2O$ in transporting other pollutants (see \cite{SherwoodEtAl}). 
The absence of links between $NO$ and $CO$ and between $NO$ and $O_3$, suggests that interactions with the other variables mediate their correlations. 
The strong effect of temperature on $H_2O$ concentration, as highlighted by jKLCV, is consistent with atmospheric physics. 
This result emphasizes the critical role of temperature in controlling humidity levels and highlights the importance of water vapour in many atmospheric processes and pollutant interactions.
The finding that temperature primarily affects $H_2O$, rather than all pollutants, aligns with the physical reality that water vapour is closely tied to thermal conditions, 
while other pollutants are more influenced by emission sources and chemical reactions in the atmosphere. 
Previous studies, such as \cite{NaumannEtAl} and \cite{SherwoodEtAl}, have explored the vertical structure of the interaction between water vapour and temperature in a non-functional framework.

\begin{figure}[!h]
  \centering
  \includegraphics[scale = .8]{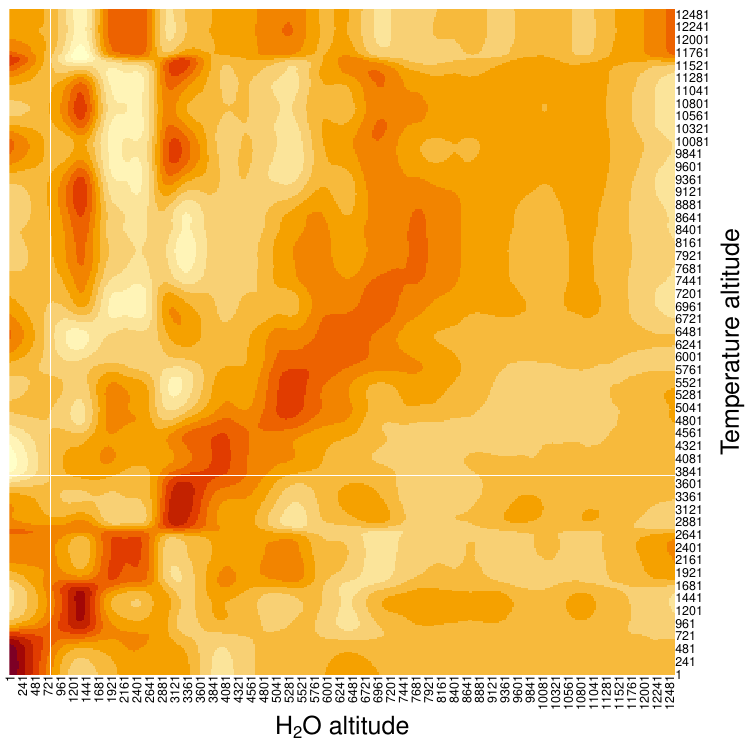}
  \caption{Value of $\widehat\beta_{H_2O}(t,s)$ for all $t,s = 1,\dots, 420$. The x-axis represents the altitude of $H_2O$, and the y-axis represents the altitude of temperature. 
  Red-coloured areas correspond to the highest regression function values ($0.00025$), and white-coloured areas correspond to the smallest values of the regression function ($-0.00006$).}
  \label{fig:beta}
\end{figure}

Figure \eqref{fig:beta} shows the estimated values of the regression function $\widehat\beta_{H_2O}(t,t^\prime)$ for all pairs $\{(t,t^\prime)$ in ${(t,t^\prime): t,t^\prime = 1, \dots, 420}\}$. 
The colour scale indicates the value of the function, with red corresponding to the highest positive values and white indicating negative values.
The red diagonal indicates that temperature has the strongest effect on $H_2O$ at the same altitude, up to approximately $8000$ meters. 
This altitude corresponds to the upper boundary of convection, a process by which warm, buoyant air rises, often leading to cloud formation in the presence of water vapour. 
Convective systems are susceptible to humidity at different heights, especially in the mid-troposphere. 
Studies such as \cite{SherwoodEtAl} emphasize that moisture in the low- to mid-troposphere (roughly $1500$ to $8000$ meters) plays a crucial role in initiating deep convection, including thunderstorms.
The colours form vertical shapes almost all over the top region of Figure \eqref{fig:beta}. 
It shows that the temperature in the upper atmosphere acts as unique block on the vapour at lower altitudes. 
The vertical shapes indicate that the effect changes over the altitude of $H_2O$. 
The lower triangle of the heatmap, which lacks red points, shows a smoother function compared to the top triangle. 
This highlights a top-down effect of temperature on water vapour, starting from layers heated by sunlight and extending downwards to the ground-level vapour concentration. 
The relationship between water vapour concentration and temperature across altitudes follows a well-known atmospheric process. Rising temperatures at lower altitudes promote evaporation, 
causing water vapour to ascend. As the vapour rises, the cooling temperature reduces its density, which is reflected by the two red points in the top-left quadrant of the heatmap. 
At certain altitudes, cooling temperatures cause water vapour to condense into clouds, resulting in concentrated cloud layers, as indicated by the white areas of the heatmap. 
Beyond this cloud layer, the air cools further, and water vapour concentrations diminish, corresponding to the positive values of the regression parameter.
As we move to higher altitudes, the colours become smoother, and the absolute values of $\beta(t,t^\prime)$ decrease, indicating a weakening influence of temperature on water vapour at these levels.

\section{Conclusion\label{sec:conclusion}}
This paper introduces a novel functional regression framework, termed \textit{functional Gaussian graphical regression model}, 
which constitutes a powerful tool for recovering relations in complex systems of functional variables, such as chemical interaction in the atmosphere. 
The primary goal of the regression analysis is to recover 
the representation of a graph with a set of undirected links and a set of directed links. 
The first reflects the interconnection of four elements, $O_3$, $NO$, $CO$ and $H_2O$, that have complex relationships in the atmosphere. Their values and interactions depend on various atmosphere conditions. 
The second set of links represents the dependences of the chemical densities in the air on the temperature.

We show that, if we assume partial separability, it is possible to analyze the conditional independence within the response functions, and the regression relations between the covariate and response functions.
We propose to estimate the model via a double-penalized joint group lasso optimization problem applied to the scores derived from expanding the joint vector of response and explanatory. 
It leads to the definition of the \textit{functional joint conditional graphical lasso} estimator.
To complete the framework, we introduce the \textit{joint Kullback-Leibler cross-validation}, a criterion tailored to select the optimal configuration for the optimization problem.
The comparison of model selection criteria reveals that jKLCV outperforms the others in terms of predictive power, particularly with small sample sizes.
The proposed framework demonstrates promising results when applied to empirical phenomena. 
From an atmospheric chemistry perspective, the jKLCV criterion provides the most plausible model, as it captures key interactions, such as those between $O_3$ and $CO$ and the peculiar role of $H_2O$, 
while not overstating the connections. 
The AIC model does not add any information to the fact that chemical interactions are intricate, something already known, while eBIC appears overly conservative. 
Thus, the jKLCV model provides a realistic representation of the interactions among elements, emphasizing the conditional nature of certain dependencies and capturing the fundamental vertical interaction between temperature and water vapour, 
which has a key role in regulating atmospheric composition. 
This approach balances model complexity and interpretability.

An interesting possible extension of the fGGR model is to relax the assumption of a multivariate Gaussian distribution of the scores using Gaussian Copula distribution, as discussed in \cite{SoleaetLi_22}.

\bibliographystyle{plainnat}
\bibliography{fcggm_ref}

\section{Appendix}\label{subsec:stcha_fun_spaces}
Let $\bm g$ and $\bm g^\prime$ be two generic set of functions in $\mathcal L_2^{p\times q}$, their inner product is defined as
 $\langle \bm g,\bm g^\prime\rangle = \int_{\mathcal{S}}\int_{\mathcal{S}} \bm g(s,t) \odot \bm g^\prime(s,t) dtds$, which leads to a $p\times q$-dimensional matrix of scalar values.
From \eqref{eq:Ymodel}, \eqref{exp.YX} and \eqref{beta_exp} we have:
\begin{eqnarray*} 
\Ev[\bm Y(t)\mid \bm X] &=& \int_{\mathcal S}\bm \beta_0(s,t)\bm X(s) ds \\
&=& \displaystyle\int_{\mathcal S} \sum\limits_{h,l=1}^\infty \Ev[\bm\gamma_{l}\bm\chi_{h}^{\top}]\Ev[\bm\chi_h\bm\chi_h^\top]^{-1}\varphi_{l}(t)\psi_h(s) \sum\limits^\infty_{h^\prime=1} \bm\chi_{h^{\prime}} \psi_{h^{\prime}}(s) ds\\
&=& \sum\limits_{h,l=1}^\infty \Ev[\bm\gamma_{l}\bm\chi_{h}^{\top}]\Ev[\bm\chi_h\bm\chi_h^\top]^{-1}\varphi_{l}(t)\sum\limits^\infty_{h^\prime=1}\bm\chi_{h^\prime} \underbrace{\displaystyle\int_{\mathcal S} \psi_h(s)\psi_{h^\prime}(s) ds}_{\small \begin{matrix} 1 \text{ if } h=h^\prime \\ 0 \text{ if } h\not=h^\prime \end{matrix} } \\
&=& \sum\limits_{h,l=1}^\infty  \Ev[\bm\gamma_{l}\bm\chi_{h}^{\top}]\Ev[\bm\chi_h\bm\chi_h^\top]^{-1}\bm\chi_h\varphi_{l}(t) 
\end{eqnarray*}

\section{Data Availability}
The data set used in the illustration is from 2020 in L2 format and is available at \\
\url{https://iagos.aeris-data.fr/download/}, 
whereas the latest version of \texttt{jcglasso} package can be found at \\
\url{https://github.com/gianluca-sottile/Hematopoiesis-network-inference-from-RT-qPCR-data/tree/main}. \\
The R script for the simulation study is available in \\
\url{https://github.com/gianluca-sottile/Functional-Gaussian-Graphical-Regression-Models}. \\
The IAGOS data analysis is available in: \\
\url{https://github.com/riifi/Functional-Gaussian-Graphical-Regression-Model}.

\section{Competing interests}
No competing interest is declared.

\section{Acknowledgments}
The authors gratefully acknowledge Rita Fici, Luigi Augugliaro and Gianluca Sottile were financially supported by the Research Projects of National Relevance - PRIN22 (CUP: B53D23009480006).
\end{document}